%% file: iclr2027_conference.tex
\documentclass{article} % For LaTeX2e
\usepackage{iclr2027_conference,times}

\input{math_commands.tex}

\usepackage{hyperref}
\hypersetup{hypertexnames=false}
\usepackage{url}

\usepackage{graphicx}
\usepackage{wrapfig}
\usepackage{subfigure}
\usepackage{tabularx}
\usepackage{booktabs}
\usepackage{multirow}
\usepackage[table,dvipsnames]{xcolor}
\usepackage{enumitem}
\usepackage{algorithm}
\usepackage{algorithmicx}
\usepackage[noend]{algpseudocode}
\usepackage{amsmath}
\usepackage{amsthm}
\usepackage{amssymb}
\usepackage{placeins}
\usepackage{tcolorbox}
\tcbuselibrary{breakable}
\usepackage{caption}
\usepackage{fontawesome5}

\definecolor{ToolGSUColor}{HTML}{70C0C7}
\definecolor{ToolIntentColor}{HTML}{FF6B6B}
\definecolor{ToolLocationColor}{HTML}{F5A623}
\definecolor{ToolFreshnessColor}{HTML}{F72D9F}
\definecolor{ToolGenderColor}{HTML}{D86BE3}
\definecolor{ToolAgeColor}{HTML}{009E73}
\definecolor{ToolPlatformColor}{HTML}{56B4E9}
\definecolor{ToolTokenColor}{HTML}{F06A00}
\definecolor{CitationGreen}{HTML}{287A4D}
\definecolor{CrossRefBlue}{HTML}{0066CC}

\definecolor{AlgorithmAccent}{HTML}{607D96}

\hypersetup{
    colorlinks=true,
    linkcolor=CrossRefBlue,
    citecolor=CitationGreen,
    urlcolor=black
}

\theoremstyle{definition}
\newtheorem{theorem}{Theorem}

\newtheorem{example}{Example}

\newcommand{\red}[1]{{\color{red}{#1}}}

\newcommand{\specialtoken}[1]{%
    \ifmmode
        \text{\ttfamily #1}%
    \else
        \texttt{#1}%
    \fi}
\newcommand{\eos}{\specialtoken{eos}}
\newcommand{\bos}{\specialtoken{bos}}

\newcommand{\stok}[2]{\specialtoken{s#1\_#2}}

\newcommand{\cttok}[2]{\textcolor{#1}{\ensuremath{t_{\text{#2}}}}}
\newcommand{\otok}[2]{\ensuremath{o_{\mathrm{#1}}\!:\!\mathrm{#2}}}
\newcommand{\cotok}[3]{\textcolor{#1}{\ensuremath{o_{\mathrm{#2}}\!:\!\mathrm{#3}}}}
\newcommand{\cls}{\specialtoken{cls}}

\title{\textit{AGR}: Agentic Generative Retrieval with Tool-Augmented Search}

\author{Ming Yang$^{1,*,\dagger}$,
Yaqi Yin$^{2,\dagger}$,
Zhibo Wang$^{2,\dagger}$,
Xiaoze Jiang$^{2,\dagger,\ddagger}$,
Jingwei Zhuo$^{3,\ddagger}$,
\\[0.6em]
\textbf{Chao Zhang}$^{2}$\textbf{,}
\textbf{Quanzhi Zhu}$^{2}$\textbf{,}
\textbf{Weiguo Zheng}$^{1}$\textbf{,}
\textbf{Liang Zeng}$^{2}$\textbf{,}
\textbf{Yunfeng Zhao}$^{2}$\textbf{,}
\\[0.6em]
\textbf{Jiaqiang Liu}$^{2}$\textbf{,}
\textbf{Kefeng Wang}$^{2}$\textbf{,}
\textbf{Yupeng Huang}$^{2}$\textbf{,}
\textbf{Kaiqiao Zhan}$^{2}$\textbf{,}
\textbf{Xuanping Li}$^{2}$
\\[0.6em]
$^{1}$Fudan University
\qquad
$^{2}$Kuaishou Technology
\qquad
$^{3}$Unaffiliated
}

\iclrfinalcopy

\begin{document}

\maketitle

\begingroup
\renewcommand{\thefootnote}{\fnsymbol{footnote}}
\footnotetext[1]{Contribution during internship at Kuaishou Technology.}
\footnotetext[2]{Equal contribution.}
\footnotetext[3]{Corresponding authors.}
\footnotetext{\faIcon[regular]{envelope}
    \{\href{yangming12@kuaishou.com}{yangming12}, \href{yinyaqi@kuaishou.com}{yinyaqi}, \href{wangzhibo07@kuaishou.com}{wangzhibo07}, \href{jiangxiaoze@kuaishou.com}{jiangxiaoze}\}@kuaishou.com, \href{zhuojw10@gmail.com}{zhuojw10@gmail.com}.
}
\endgroup

\vspace{-5mm}
\begin{abstract}
\vspace{-3mm}

Large-scale search and recommendation systems derive much of their retrieval quality from rich, heterogeneous features, yet existing generative retrieval (GR) methods struggle to exploit such feature ecosystems efficiently. Static feature injection either encodes all available signals or places them according to predetermined schemes, increasing serving cost and introducing irrelevant or conflicting evidence. More fundamentally, existing methods cannot adaptively determine when a feature is needed, which feature to select, or where it should influence hierarchical Semantic ID (SID) generation. We propose \textit{Agentic Generative Retrieval (AGR)}, a tool-augmented paradigm that interleaves SID generation with on-demand feature acquisition. Conditioned on the request and decoded SID prefix, \textit{AGR} decides whether to invoke a tool, which feature to acquire, and where its observation should influence subsequent decoding, thereby steering generation toward feature-consistent SID branches. To supervise this behavior, we construct tool-augmented trajectories based on feature applicability and information gain along the SID hierarchy, and train the model through SID pretraining followed by tool-augmented fine-tuning. Experiments on an industrial dataset and three public benchmarks, together with an online A/B test, consistently demonstrate the effectiveness of \textit{AGR}.

\end{abstract}
\input{sections/introduction}

\input{sections/problem}
\input{sections/analysis}
\input{sections/method}
\input{sections/experiments}
\input{sections/limitations}
\input{sections/conclusion}

\clearpage

\bibliography{iclr2027_conference}
\bibliographystyle{iclr2027_conference}

\clearpage
\appendix
\input{sections/appendix}

\end{document}

%% file: math_commands.tex
\usepackage{amsmath,amsfonts,bm}

\def\eqref#1{equation~\ref{#1}}
\def\1{\bm{1}}

\DeclareMathAlphabet{\mathsfit}{\encodingdefault}{\sfdefault}{m}{sl}
\SetMathAlphabet{\mathsfit}{bold}{\encodingdefault}{\sfdefault}{bx}{n}

\DeclareMathOperator*{\argmax}{arg\,max}

%% file: sections/introduction.tex
\vspace{-3mm}
\section{Introduction}
\label{sec:introduction}
\vspace{-2mm}

Generative Retrieval (GR) casts retrieval as autoregressive identifier generation \citep{tay2022transformermemorydifferentiablesearch,wang2023neuralcorpusindexerdocument,bevilacqua2022autoregressivesearchenginesgenerating}. Following TIGER~\citep{TIGER}, items are commonly represented by hierarchical Semantic IDs (SIDs) learned through residual quantization. This paradigm has also been successfully adopted in large-scale industrial scenarios \citep{Onerec, Onesearch, Unisearch, Onereason}.

Deploying GR in industrial systems, however, exposes a fundamental mismatch between request-conditioned SID generation and the feature-intensive nature of such systems~\citep{HSTU}. 
Large-scale retrieval depends heavily on heterogeneous signals, such as user profiles, behavioral histories, and freshness cues~\citep{SubtleQuery}. 
Yet existing GR methods scale poorly with the number of features (Figure~\ref{fig:sequence-cost}), typically appending all signals to the input or injecting them according to a predetermined scheme, making it difficult to jointly maintain retrieval quality and serving efficiency~\citep{GEMs}.
Specifically, encoding all features indiscriminately inflates serving cost and introduces irrelevant or conflicting evidence, as feature utility is inherently request-dependent: a feature indispensable for one request may be pure noise for another~\citep{RecThinker}. This raises the question of \textbf{when a feature should be incorporated}.
Meanwhile, injecting features at a fixed scheme cannot adapt to the decoding state, but residual quantization induces a hierarchical SID space where different depths resolve distinctions at different granularities. Therefore, a feature highly informative at one level may be irrelevant at others. This raises the question of \textbf{where a feature should influence generation}.
Addressing both questions is crucial for efficient and scalable GR deployment in industrial search, yet effective feature injection remains largely overlooked.

To this end,
we propose \textbf{\underline{A}gentic \underline{G}enerative \underline{R}etrieval (\textbf{AGR})}, which reformulates generative retrieval as an interleaved process of SID generation and feature acquisition.
\textit{AGR} encapsulates heterogeneous features as tools, each uniquely identified by a tool token.
Conditioned on the current decoding state, the model decides whether to emit a tool token and, if so, which tool to invoke.
The acquired feature values are then represented as observation tokens that reshape the probability distribution over subsequent SID tokens, steering generation toward feature-consistent paths.
As features are introduced only when the decoding state warrants them, \textit{AGR} reduces feature noise and naturally aligns feature granularity with the coarse-to-fine progression of SID generation.
Through this adaptive tool use, \textit{AGR} shifts generative retrieval from passive SID generation to agentic decision-making over both items and features.

\noindent \textbf{Contributions.} Our contributions are shown as follows:
\begin{itemize}[leftmargin=*]

\item 
Driven by the industrial need to exploit rich, heterogeneous feature ecosystems in GR, we systematically reveal and quantify two fundamental limitations of static feature injection in GR: it is neither request-adaptive (\emph{when} to inject) nor level-adaptive (\emph{where} to inject). We further argue that features should be acquired on demand and injected at the right SID depth.

\item We propose \emph{Agentic Generative Retrieval (\textit{AGR})}, which encapsulates feature injection as tool invocation and thereby unifies adaptive feature selection with dynamic placement into a single decoding process, paving the way for exploiting rich features in real-world retrieval systems.

\item We derive principled criteria for choosing which tool to invoke and where it is most informative along the SID hierarchy, so that each tool call steers decoding toward feature-consistent branches, and instantiate them in a two-stage pipeline of SID pretraining and tool-augmented fine-tuning.

\item We evaluate \textit{AGR} on Kuaishou Search system and three widely used public benchmarks. \textit{AGR} improves recall by up to +14.04\% over baseline and yields consistent online A/B gains.
Extensive experimental analyses further explain the gains achieved by \textit{AGR}.
\end{itemize}

%% file: sections/problem.tex
\section{Problem Definition}
\label{sec:preliminaries}
\vspace{-2mm}

For reference, the notation used throughout this paper is summarized in
Table~\ref{tab:notation}.

\paragraph{Generative Retrieval.}
Given a request containing query $q$ and model-visible user context $u_{\mathrm{vis}}$, generative retrieval aims to retrieve $K$ items from a corpus $\mathcal{X}$ by directly generating item identifiers. 
Each item $x\in\mathcal X$ is represented by a hierarchical SID sequence $\mathbf s=[s_{1},\ldots,s_{L}]$, where $s_{i}\in\mathcal S_i$ denotes the token at level $i$ and $\mathbf{s}_{\leqslant i} = [s_{1}, \ldots, s_{i}]$ denotes its prefix.
The retriever generates the sequence autoregressively by modeling
$\mathbb{P}_\theta(s_{i+1} \mid q, u_{\mathrm{vis}}, \mathbf{s}_{\leqslant i})$, and scores each prefix by accumulating token-level log-probabilities 
$
\pi_\theta(\mathbf{s}_{\leqslant i}) = \sum_{j=1}^{i}\log \mathbb{P}_\theta\left(s_j\mid q, u_{\mathrm{vis}}, \mathbf{s}_{\leqslant j-1}\right)
$.
During inference, a prefix trie $\mathrm{Tr}$ over valid SIDs constrains beam search to valid paths only.
After $L$ decoding steps, the top-$K$ completed sequences are mapped to the corresponding items as results~\citep{decao2021autoregressiveentityretrieval}.

\paragraph{Feature Injection.}
Beyond the visible request context $q$ and $u_{\mathrm{vis}}$, 
practical GR systems often rely on heterogeneous auxiliary features, including long-term behaviors, user attributes, and other auxiliary signals stored in external services. 
We denote the candidate feature set as $\mathcal{F}$. 
Incorporating a subset $\mathcal{F}' \subseteq \mathcal{F}$ into decoding is referred to as \emph{feature injection}, which falls into two categories:
\begin{itemize}[leftmargin=*]
\item \textbf{Static injection} commits to a fixed subset $\mathcal{F}_{\mathrm{static}} \subseteq \mathcal{F}$ before decoding and reuses it at every step, so the same features influence the entire SID generation:
% \vspace{-3mm}
\begin{equation}
\pi_\theta(\mathbf{s})
=
\sum_{i=1}^{L} \log
\mathbb{P}_\theta(s_{i}\mid q, u_{\mathrm{vis}}, \mathbf{s}_{\leqslant i-1}, \mathcal{F}_{\mathrm{static}}).
\label{eq:static-injection}
\end{equation}
% \vspace{-3mm}
\item \textbf{Dynamic injection} selects a different subset at each step conditioned on the decoded prefix $\mathbf{s}_{x,<i}$, 
allowing both the identity and the number of injected features to vary across decoding depths:
% \vspace{-3mm}
\begin{equation}
\pi_\theta(\mathbf{s})
=
\sum_{i=1}^{L} \log
\mathbb{P}_\theta(s_{i}\mid q, u_{\mathrm{vis}}, \mathbf{s}_{\leqslant i-1}, \mathcal{F}_i),
\quad \text{ where }
\mathcal{F}_i \subseteq \mathcal{F},\;
\mathcal{F}_i = \sigma(q, u_{\mathrm{vis}}, \mathbf{s}_{\leqslant i-1}),
\label{eq:dynamic-injection}
\end{equation}
% \vspace{-3mm}
where $\sigma(\cdot)$ is a prefix-aware feature selection policy.
Static injection is recovered as the special case
$\mathcal{F}_i \equiv \mathcal{F}_{\mathrm{stat}}$ for all $i$.
\end{itemize}

%% file: sections/analysis.tex
\vspace{-3mm}
\section{Limitations of Static Feature Injection}
\label{sec:analysis}
\vspace{-2mm}
Having defined static and dynamic feature injection in GR, we now analyze why the static scheme adopted by most existing GR methods is insufficient~\citep{TIGER, GEMs, RecThinker, OneSearch-V2}. 
Motivated by the observed cost--gain imbalance of static injection at scale~\citep{GEMs}, we compare a shared feature-free baseline with feature-specific static-injection models, where each model statically encodes one feature type into the input.

\begin{wrapfigure}{r}{0.45\textwidth}
    % \vspace{-2pt}
    \centering
    \includegraphics[width=0.43\textwidth]{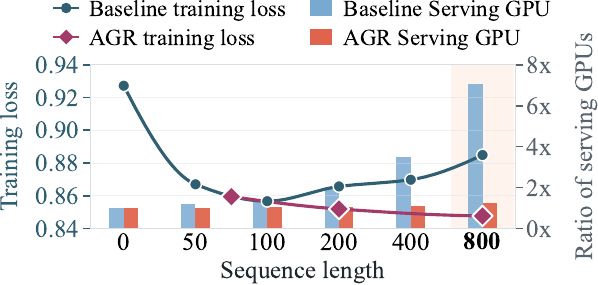}
    \vspace{-8pt}
    \caption{
    % \small 
    Feature injection efficiency on training loss and serving cost.}
    \label{fig:sequence-cost}
    \vspace{-14pt}
\end{wrapfigure}

We begin with an \textbf{empirical observation}.
As shown in Figure~\ref{fig:sequence-cost}, increasing the amount of features injected statically, instantiated by user behavior sequences, leads to diminishing or even negative improvements for training, while increasing serving cost.
This suggests that indiscriminate, static feature injection is neither consistently beneficial nor cost-efficient. 
Moving beyond static injection, dynamic injection raises two key challenges: \emph{when} a feature should be incorporated across requests, and \emph{where} in the SID hierarchy it should take effect.

\begin{wrapfigure}{r}{0.45\textwidth}
    \vspace{-10pt}
    \centering
    \includegraphics[width=0.43\textwidth]{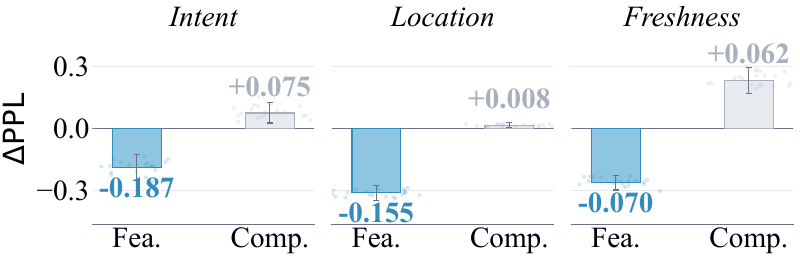}
    \vspace{-5pt}
    \caption{
    % \small 
    Static injection on feature-aligned (Fea.) and complementary (Comp.) subsets.
    }
    \label{fig:when-analysis}
    \vspace{-3mm}
\end{wrapfigure}

For \textbf{the first challenge}, 
static injection assumes an injected feature is universally useful, 
rather than request-dependent.
To diagnose this issue, we construct two test sets:
feature-consistent set Fea., where the request and target item share the same feature value, and feature-conflicting set Comp., where their values differ.
We measure the injection effect by the perplexity~\citep{PPL}:
% change on the golden SIDs $[s_1,\ldots,s_L]$:
\vspace{-1mm}
\begin{equation}
    % \small
    % \quad
\Delta\mathrm{PPL}=\mathrm{PPL}_{\mathrm{static}}-\mathrm{PPL}_{\mathrm{w/o\ feature}}, \quad \mathrm{PPL}=\exp\left(-\frac{1}{L}\sum_{i=1}^{L}\log \mathbb{P}(s_i\mid q, u_{\mathrm{vis}}, \mathbf{s}_{\leqslant i-1})\right).
\label{eq:ppl}
\end{equation}
% \vspace{-1mm}
As shown in Figure~\ref{fig:when-analysis}, the static injection model lowers $\mathrm{PPL}$ in Fea. set but raises it in Comp. set, confirming that feature utility is request-dependent.
Therefore, efficient feature injection should be request-adaptive, using features only when beneficial.

\begin{wraptable}{r}{0.45\textwidth}
    % \vspace{-12pt}
    \centering
    \scriptsize
    \setlength{\tabcolsep}{6pt}
    \renewcommand{\arraystretch}{1.2}
        \caption{
        % \small 
        Golden-token probability ratios under static feature injection.}
    \label{tab:static-where}
    \begin{tabular}{clrr}
        \toprule
        Illustration & Feature & $r_{i=i^*}$ & $r_{i\neq i^*}$ \\
        \midrule
        \multirow{3}{*}{\includegraphics[width=0.078\textwidth]{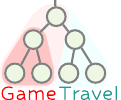}}
               & Intent & 1.032 & 1.040 \\
               & Location & 1.003 & 1.001 \\
               & Freshness & 1.008 & 1.010 \\
        \bottomrule
    \end{tabular}
\end{wraptable}

For \textbf{the second challenge}, 
static injection exposes all decoding steps to the same features.
However, RQ-based SIDs typically exhibit coarse-to-fine semantics,
suggesting that different steps may require different features.
To verify this, we identify, for each feature, the depth $i^*$ where injecting it would most improve next-token prediction (detailed in Section~\ref{sec:training}), and then measure whether static injection actually concentrates its effect at the depth $i^*$. Concretely, we compare the golden-token probability ratio at each step:
\vspace{-2mm}
\begin{equation}
% \small
    r_i =
    \frac{\mathbb P_{\mathrm{static}}(s_{i+1} \mid q, u_{\mathrm{vis}}, \mathbf{s}_{\leqslant i})}{\mathbb P_{\mathrm{w/o\ feature}}(s_{i+1} \mid q, u_{\mathrm{vis}}, \mathbf{s}_{\leqslant i})} \, ,
\label{eq:prob-gain-ratio}
\end{equation}

where $\mathbb{P}_{\mathrm{static}}$ and $\mathbb{P}_{\mathrm{w/o\ feature}}$ denote the golden-token probability produced by models with features statically injected into the model input and without any encoded features, respectively.
As shown in Table~\ref{tab:static-where}, the probability ratio $r_{i^*}$ is not consistently higher than at other levels. This confirms that static injection spreads its influence uniformly rather than concentrating it where the feature is most needed.
Efficient feature injection therefore requires a level-adaptive mechanism that activates each feature only at the decoding depths where it is expected to be informative.

%% file: sections/method.tex
% \vspace{-3mm}
\section{Methodology}
\label{sec:method}
% \vspace{-2mm}

To enable on-demand feature interaction, we propose \textit{\underline{A}gentic \underline{G}enerative \underline{R}etrieval (AGR)}. Section~\ref{sec:agr-method} introduces the inference framework and tool implementation; Section~\ref{sec:training} describes how data are constructed and used to operationalize this tool-augmented generative retrieval paradigm.
% \vspace{-3mm}
\subsection{Agentic Generative Retrieval}
\label{sec:agr-method}
% \vspace{-2mm}
\begin{figure}[ht]
    \centering
    \includegraphics[width=\linewidth]{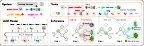}
    % \vspace{-7mm}
    \caption{
        % \small
Overview of \textit{AGR}:  
(1) \textit{AGR Tool} extends the vocabulary with tool and observation tokens, mapping tool calls to feature values; CF. are categorical features; GSU (General Search Unit) are user behavioral sequences.
(2) \textit{AGR Model} performs SID generation by predicting tool tokens and adaptively injecting external features via observation tokens (QU: query text and user features).
(3) \textit{AGR Inference} enables model-tool interaction for each beam candidate during decoding, forming tool-adaptive retrieval.
    }
    \label{fig:overview}
\end{figure}

% \vspace{-4mm}
\paragraph{\textit{AGR} Framework.}
To provide a flexible and extensible interface for feature interaction, we model tool calls and their results as tokens.
Specifically, each tool token $t \in \mathcal{T}$ triggers a function call $f_t(q, u, \mathbf{s}_{\leqslant i})$, whose result is represented as an observation token $o_t \in \mathcal{O}$.
By extending the vocabulary to
$\mathcal{V} = \mathcal{S} \cup \mathcal{T} \cup \mathcal{O}$, the model naturally unifies SID generation and tool invocation into a single autoregressive process.
As shown in Figure~\ref{fig:overview}, given the request \( q \), user \( u_{\mathrm{vis}} \), decoded prefix $\mathbf{s}_{\leqslant i}$, 
and prior observations, the model predicts either the next SID token $s_{i+1}\in\mathcal{S}$ or a tool token $t\in\mathcal{T}$ (e.g. \cttok{ToolIntentColor}{I} or \cttok{ToolGSUColor}{G}). 
Specifically, tool tokens serve solely as decision triggers and are not fed back into the autoregressive context;
only the returned observation token (e.g., $\textcolor{ToolIntentColor}{o_{\mathrm{I}}}$ or $\textcolor{ToolGSUColor}{o_{\mathrm{G}}}$) is injected, 
with distinct positional encodings to differentiate it from $\mathcal{S}$.
As a result, the decoder consumes a single token per step,
yet still allowing observations to prune irrelevant branches during generation
(e.g., the ``Movie'' branch in Figure~\ref{fig:overview}).
For more details, refer to Algorithm~\ref{alg:tool-beam-search} and Example~\ref{ex:agr-workflow}.

\begin{figure*}[htbp]
    \centering
    \vspace{-5mm}
    \includegraphics[width=0.8\textwidth]{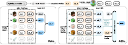}
    \vspace{-2mm}
    \caption{
        Two variants of the sequential feature tool.
    }
    \label{fig:gsu-tool}
    \vspace{-3mm}
\end{figure*}

\paragraph{Tool Implementations.}
\label{pgh:tool-impl}
Based on the type of observation returned, tools fall into two categories:
\emph{structured feature tools}, 
and \emph{sequential feature tools}:
\begin{enumerate}[label=(\alph*), leftmargin=*]

\item \emph{Structured feature tools} return categorical values and can be directly encoded into a single learnable observation token —
e.g., a game intent maps to $\cotok{ToolIntentColor}{I}{Game}$, a Beijing location to $\cotok{ToolLocationColor}{L}{Beijing}$, and a high freshness requirement to $\cotok{ToolFreshnessColor}{F}{High}$.
Due to the hierarchical structure of RQ,
items with similar features naturally share SID prefixes;
for example, videos about games tend to cluster under the same prefixes~\citep{TIGER}.
Thus, these learnable observation tokens could serve as informative signals during decoding.

\item \emph{Sequential feature tools} return user behavioral sequences (e.g., General Search Unit, GSU~\citep{GEMs}), which need to be encoded into a single observation token $\textcolor{ToolGSUColor}{o_{\mathrm{G}}}$. 
Various methods could accomplish this~\citep{GLASS}; we adopt two complementary approaches:
\emph{Vote}, a parameter-free approach that is efficient for training and inference, and \emph{Attn}, which offers stronger user preference modeling capacity.
As shown in Figure~\ref{fig:gsu-tool}, 
\emph{Vote} performs prefix matching:
it selects items whose SID prefix matches the decoded one, then picks the most frequent next SID token (i.e., {\stok{3}{7}}) via majority voting.
\emph{Attn} instead operates in the embedding space:
it computes the accumulated residual representation of the decoded prefix (e.g., the sum of the embeddings of \stok{1}{1} and \stok{2}{3}),
retrieves the Top-$E$ ($E=5$ for evaluation) most similar items from the sequence,
and encodes their next SID token representations into a single embedding via multi-head attention.
Unless otherwise stated, we use the \emph{Vote} variant (Table~\ref{tab:fusion-strategy}).

\end{enumerate}

\vspace{-3mm}
\subsection{Tool-augmented \textit{AGR} Fine-tuning}
\label{sec:training}
\vspace{-2mm}
\paragraph{Data Construction.} 
Having incorporated tool calls and observations into the token vocabulary (Section~\ref{pgh:tool-impl}),
we now describe how to construct training data that teaches the model to invoke tools during decoding. 
Concretely, for each positive sample $x^+$ with target SIDs $\mathbf{s}^+=[s_1,\ldots,s_L]$, 
we augment the sequence by inserting a tool token $t$ at depth $i^*$ and its corresponding 
observation $o_t$ into the shifted input, 
so that the model predicts $[s_1,\ldots,s_{i^*},t,s_{i^*+1},\ldots,s_L,\eos]$ 
from $[\bos,s_1,\ldots,s_{i^*},o_t,s_{i^*+1},\ldots,s_L]$. 
As outlined in Section \ref{sec:analysis}, this construction requires answering two questions: 
\textit{when} --- which samples should be augmented with which tools, 
and \textit{where} --- at which decoding level $1\leqslant i^* < L$ the tool should be invoked.

\begin{itemize}[leftmargin=*]
\item \textbf{``When'' is a feature applicable?}
\label{sec:tool-applicability} 
From the perspective of individual samples, tools are not universally
applicable across all scenarios. Leveraging available supervised data, we
identify when a tool is applicable and construct
\(\mathcal{P}(x^+)\), the set of applicable tool-position pairs
\((t,i)\) for each positive sample \(x^+\).
\begin{enumerate}[label=(\alph*)]
    \item 
    For a \textit{structured feature tool} $t$, $o_t = f_t(q,u)$ extracts the feature from the request side (e.g., query and all user context) to produce observation $o_t$, and $g_t(x^+)$ extracts the same feature from the positive sample $x^+$.
    To ensure both relevance and non-redundancy, we retain $t$ only when the two extractions agree, and the feature is not already explicit in the  context:
    \begin{equation}
        % \small
        \mathcal P_{\mathrm{str}}(x^+)
    =\{(t,i)\mid o_t=g_t(x^+) \land\ o_t\text{ absent from model inputs},\ 1\leqslant i<L\}.
    \label{eq:Pstr}
    \end{equation}

\item
For a \textit{sequential feature tool} $t$, 
$o_{t,i}=f_t(q, u,\mathbf{s}^{+}_{\leqslant i})$
derives its observation from user behavior sequences,
while $g_t(x^+,\mathbf{s}^+_{\leqslant i}) = s^+_{i+1}$ directly extracts the golden next SID.
We instantiate $f_t$ with the \emph{Vote} variant $f_\mathrm{vote}$.
Similarly, we retain $t$ only when the two extractions agree: 

\begin{equation}
% \small
\mathcal{P}_{\mathrm{seq}}(x^+)
=
\left\{
(t,i)
\mid
o_{t,i}=s^{+}_{i+1},
\ 1\leqslant i<L
\right\}.
\label{eq:Pseq}
\end{equation}

To make the model robust to ineffective tool calls, we also sample incorrect predictions, maintaining a voting hit rate of $\tau$ to control the hit-to-miss ratio.

\end{enumerate}

\item \textbf{``Where'' to insert the feature?}
\label{sec:observation-insertion}
Having identified the samples amenable to tool insertion, we now determine the optimal depth $i^*$ to insert tools.
The key intuition is that a feature should be injected at the depth where it best disambiguates the next SID token.
We formalize this via $\rho_i$, 
which reflects how discriminative the feature is at that prefix.
The depth where $\rho$ changes most sharply is selected as the most valuable position for feature injection:

\vspace{-2mm}

\begin{equation}
% \small
i_t^*(x^+)
=
\operatorname*{arg\,max}_{\substack{i}}
\left[
\rho_{i+1}(o_t)-\rho_i(o_t)
\right] \quad \text{s.t.}\  (t,i)\in\mathcal{P}(x^+),
\label{eq:observation-position}
\end{equation}

where \(\mathcal{P}=\mathcal{P}_{\mathrm{str}}\) for structured feature
tools and \(\mathcal{P}=\mathcal{P}_{\mathrm{seq}}\) for sequential
feature tools. We define \(\rho_i\) for each tool type as follows:

\begin{enumerate}[label=(\alph*)]
\item 
For a \textit{structured feature tool}, the observation is a discrete attribute of items.
To identify where injecting this feature narrows the candidate set most, therefore best disambiguates the next SID token, we compute the log-proportion of items under prefix $\mathbf{s}^+_{\leqslant i}$ that share the same attribute value over the corpus $\mathcal{X}$:
\begin{align}
\rho_i(o_t)
= \log \mathbb{P}_{x' \sim \mathcal{X}}\left(g_t(x') = o_t \mid \mathbf{s}^+_{\leqslant i}\right).
\label{eq:structured-feature-concentration}
\end{align}
The sharpest increase in $\rho_i$ identifies the depth 
where the feature becomes most discriminative,
therefore provides the strongest signal for predicting the next SID.

\item 
For a \textit{sequential feature tool}, the observation is the next SID token voted from behavior sequences.
To measure where this observation narrows the candidate set most, we compute the log-proportion of items under prefix $\mathbf{s}^+_{\leqslant i}$ that share the same next token over $\mathcal{D}_{\mathrm{train}}$:
\vspace{-2mm}
\begin{align}
\rho_i(o_{t,i})
=
\log \mathbb{P}_{x'\sim\mathcal{D}_{\mathrm{train}}}
\left(
g_t(x',\mathbf{s}'_{\leqslant i})=o_{t,i}
\mid
\mathbf{s}_{\leqslant i}^{+}
\right).
\label{eq:sequential-feature-concentration}
\end{align}
By Theorem~\ref{thm:equivalent-criteria-for-sequential-feature-placement},
maximizing the increase in $\rho_i$ is equivalent to maximizing the
model's prediction gain (i.e.,
$r_i=\mathbb{P}_{\theta}/\mathbb{P}_{\text{w/o features}}$, analogous
to Eq.~\ref{eq:prob-gain-ratio}); both are further equivalent to
minimizing $\rho_i$ itself. 
Accordingly, for sequential feature placement, we typically adopt the
equivalent criterion \(i^* = \arg\min_i \rho_i(o_{t,i})\).
Intuitively, a lower $\rho_i$ indicates that the golden SID is rarer
among items sharing the same prefix, making the observation more
informative for identifying the user's fine-grained or niche
preferences.
\end{enumerate}

The ``when'' and ``where'' decisions are illustrated in Example~\ref{ex:training-data-construction}.
\end{itemize}

\vspace{-2mm}
\paragraph{Two-Stage Training.}
To decouple semantic learning from tool integration for better extensibility,
\textit{AGR} training consists of two stages. 
\textbf{The first stage} leverages GR (SID-only) data to learn SID generation following traditional GR methods~\citep{Unisearch}. 
\textbf{The second stage} then introduces \textit{AGR} data, where the model learns to integrate tool calls into generation.
The loss now covers both SID tokens and tool tokens.
    In practice, since tool tokens are unseen during the first stage, their 
    token-level losses are initially much higher than those of SID tokens, risking destabilization of already-learned representations.
    To mitigate this imbalance, we scale the loss of each tool token by a factor $\alpha_k$ that adapts per training step.
    Formally, let $\mathbf v_{\mathrm{out}}(x)=[v_{1},\ldots,v_{|\mathbf v_\mathrm{out}|}]$ denote the decoder output sequence for the given sample $x$.
    For mini-batch $\mathcal B_k$ at step $k$, we measure the prediction difficulty gap between tool and SID tokens:
$
% \small
\beta_k=
\frac{
\mathbb E_{(x,j)\sim\mathcal B_k}
[\log \mathbb{P}_\theta(v_{j}\mid q, u_{\mathrm{vis}}, \mathbf{s}_{\leqslant j-1})\mid v_{j}\notin\mathcal T]
}{
\mathbb E_{(x,j)\sim\mathcal B_k}
[\log \mathbb{P}_\theta(v_{j}\mid q, u_{\mathrm{vis}}, \mathbf{s}_{\leqslant j-1})\mid v_{j}\in\mathcal T]
}
$ 
and clip it to [$\alpha_{\min}$, $\alpha_{\max}$] for stability 
$
\alpha_k = \mathbf{1}[v_{j}\in\mathcal{T}]\cdot\operatorname{clip}(\beta_k,\alpha_{\min},\alpha_{\max}) + \mathbf{1}[v_{j}\notin\mathcal{T}].
$
The final objective is then delivered as:
\begin{equation}
        \mathcal L_{\mathrm{AGR}}=-\frac{1}{|\mathcal{B}_k|}\sum_{x\in\mathcal \mathcal{B}_k}\frac{1}{|\mathbf v_\mathrm{out}|}\sum_{j=1}^{|\mathbf v_\mathrm{out}|}\alpha_k \cdot \log \mathbb{P}_\theta(v_{j}\mid q, u_{\mathrm{vis}},\mathbf{s}_{\leqslant j-1}).
\label{eq:agr_loss}
\end{equation}

%% file: sections/experiments.tex
% \vspace{-10mm}
\section{Experiments}
\label{sec:experiments}
% \vspace{-2mm}
\subsection{Experimental Setup}
\label{sec:experimental-setup}
% \vspace{-2mm}
\paragraph{Datasets.}
We evaluate \textit{AGR} on four datasets: Industry, Qilin, JDSearch, and KuaiSAR (large). 
Industry is derived from real-world user behavior logs collected from Kuaishou Search, with approximately 300 million records per day; we use June 14--20 for the first training stage, June 21--22 for the second training stage, and June 23 for testing. 
Qilin is a multimodal retrieval benchmark with app-level user sessions~\citep{chen2025qilin}; JDSearch focuses on personalized product-search benchmark with real queries and user--item interactions~\citep{liu2023jdsearch}; and KuaiSAR provides a unified search-and-recommendation benchmark~\citep{sun2023kuaisar}. 
Further details are provided in Table~\ref{tab:dataset-overview}.

\vspace{-2mm}
\paragraph{\textit{AGR} Implementations.}
We adopt a 12-layer encoder-decoder BART~\citep{BART} as the backbone ($\approx$ 574M parameters, including embeddings for approximately 19,000 predefined SID codewords) and construct hierarchical SIDs via RQ-KMeans~\citep{TIGER}.
Dataset-specific SID configurations and tool sets are listed in Table~\ref{tab:tool-configs}; inference follows Algorithm~\ref{alg:tool-beam-search}.
On the large-scale Industry dataset, we train the model sequentially in two stages, with each stage running for one epoch over its corresponding aggregated training data.
For the smaller public datasets, we fine-tune with each method for 10 epochs until convergence. 
For fair comparison, within each dataset, all methods share the same text-pretrained backbone, SID codebooks, training data, learning rate, and hardware configuration. 
Additional details are provided in Appendix~\ref{app:implementation-details}.

\vspace{-2mm}
\paragraph{Evaluation Metrics.}
We adopt Recall@$K$ and Mean Reciprocal Rank (MRR@$K$) with $K\in\{32,50,64\}$ ($K=50$ for online A/B test).
Given $N$ test requests, for $i$-th query, let $\mathcal G_i$ denote the ground-truth item set, $\mathcal R_i$ the top-$K$ retrieved items, and $r_i$ the rank of the first ground-truth item ($r_i=\infty$ if none appears in the top-$K$).
Formally, we have:
\begin{equation}
% \small
    \operatorname{Recall@}K=\frac{1}{N}\sum_{i=1}^{N}
    \frac{|\mathcal G_i\cap\mathcal R_i|}{|\mathcal G_i|}
\quad
\text{and}
\quad
    \operatorname{MRR@}K=\frac{1}{N}\sum_{i=1}^{N}
    \frac{\mathbb I[r_i\leqslant K]}{r_i}.
\end{equation}

% \vspace{-3mm}
\subsection{Main Results}
\label{sec:main-results}
\vspace{-2mm}
As shown in Table~\ref{tab:main-results},
\textbf{\textit{AGR} consistently outperforms all baselines across all
datasets and evaluation metrics.}
We compare \textit{AGR} with five generative retrieval baselines:
TIGER~\citep{TIGER} for standard GR baseline; TIGER Distill Teacher,
which first trains a teacher model with up to 400 GSU tokens
(Table~\ref{tab:sequential-encoding}) and then distills its predictions
into TIGER using an auxiliary distillation loss with a weight of $0.1$;
UniSearch~\citep{Unisearch} for unified short-video retrieval; and
GLASS~\citep{GLASS} and GEMs~\citep{GEMs} for long-history modeling.
\textit{AGR} achieves the best Recall and MRR at all cutoffs, with
relative improvements over TIGER ranging from 6.4\% to 14.0\% in Recall
and from 5.2\% to 19.7\% in MRR. Its advantages over TIGER Distill
Teacher and UniSearch demonstrate the benefit of dynamically acquiring
request-specific evidence during decoding, a capability absent from
these methods. Its advantages over GLASS and GEMs further demonstrate
that adaptive tool integration exploits long-term behavioral features
more effectively than direct sequence modeling.

\input{sections/main_results_table}

% \clearpage
% \vspace{-8mm}
\subsection{Ablation Study}
\label{sec:ablation}
% \vspace{-2mm}

Our ablation and analysis primarily focus on the Industry dataset, as it supports more diverse tool construction than existing public benchmarks. Additional ablations and analyses are in Section~\ref{app:ablation}.

\vspace{-2mm}
\par\WFclear
\begin{wraptable}{r}{0.45\textwidth}
    \vspace{-12pt}
    \caption{\small 
    % \red{
    Varying feature interaction strategies.
    % }
    }
    \label{tab:feature-interaction}
    \vspace{-2mm}
    \centering
    \scriptsize
    \setlength{\tabcolsep}{12pt}
    \renewcommand{\arraystretch}{0.92}
    \begin{tabular}{@{}lr@{}}
        \toprule
        Strategies & Recall@50 \\
        \midrule
        (a) Static feature injection & 0.3159 \\
        (b) Query-guided gating & 0.3161 \\
        (c) Fixed-position tool injection & 0.3174 \\
        \textbf{(d)} \textbf{\textit{AGR} (ours)} & \textbf{0.3264} \\
        \bottomrule
    \end{tabular}
    \vspace{-6pt}
\end{wraptable}

\paragraph{Adaptive feature injection outperforms predetermined strategies.}
A defining principle of \textit{AGR} is that feature injection should be driven by the model's generation state rather than predetermined.
We ablate this across four strategies over a feature set including intent, location, freshness and GSU sequence.  
\emph{Static feature injection} feeds all features into the encoder without selection or weighting. 
\emph{Query-guided gating} uses a lightweight MLP, whose parameter count is substantially smaller ($\approx$4,100 parameters) than that of the backbone, to learn query-dependent feature weights.
\emph{Fixed-position tool injection} retrieves features via tool calls, but at a predetermined position. 
\textit{AGR} uses the same tool-call mechanism but injects adaptively. 
As shown in Table~\ref{tab:feature-interaction}, 
query-guided gating and fixed-position injection both outperform static injection, yet \textit{AGR} achieves the best Recall@50, 
confirming the effectiveness of adaptive injection.

\FloatBarrier

\begin{wraptable}{r}{0.5\textwidth}
    \vspace{-10pt}
    \caption{
    Comparison of direct GSU sequence encoding and dynamic acquisition in \textit{AGR}.
    Training time is measured under identical batch size and configuration. Serving cost is the number of GPUs required per 1K requests under the same offered load and GPU configuration.}
    \vspace{-3pt}
    \label{tab:sequential-encoding}
    \centering
    \scriptsize
    \setlength{\tabcolsep}{2pt}
    \renewcommand{\arraystretch}{0.92}
    \begin{tabular}{@{}lccc@{}}
\toprule
\multirow{2}{*}{Method}
& \multirow{2}{*}{Recall@50}
& Train Time
& Serving Cost \\
& & (sec/step) & (GPUs/1K req.) \\
\midrule
Baseline
& 0.2863 & 1.163 & 2.38 \\

+ 50 relevant GSU tokens
& 0.3195 & 2.413 & 2.70 \\

+ 100 relevant GSU tokens
& 0.3202 & 3.671 & 3.03 \\

+ 400 relevant GSU tokens
& 0.3241 & 10.794 & 8.33 \\

\textbf{\textit{AGR} w/ GSU tool only}
& \textbf{0.3237} & \textbf{1.969} & \textbf{4.76} \\
\bottomrule
\end{tabular}
    \vspace{-8pt}
\end{wraptable}

\vspace{-2mm}
\paragraph{Sequential feature tools match long sequence encoding at a fraction of the cost.}
% \paragraph{Sequential feature tools nearly match long-sequence encoding at substantially lower computational cost.}
Tool-based retrieval compresses sequential features into discrete observations to keep serving cost manageable. We quantify this trade-off by comparing \textit{AGR} with directly appending up to 400 GSU tokens to the input.
As shown in Table~\ref{tab:sequential-encoding}, direct encoding improves Recall@50 from 0.2863 to 0.3241, but at steep cost: training time rises from 1.163 to 10.794 seconds per step, and serving cost rises from 2.38 to 8.33 GPUs per 1K requests. 
In contrast, \textit{AGR} achieves comparable Recall@50 of 0.3237, 
while reducing training time and serving cost by 81.8\% and 42.9\%. 
Although \textit{AGR} roughly doubles the serving cost over the no-feature baseline, it remains far more efficient than direct encoding.
% We also observe a similar trade-off for structured features (Appendix~\ref{app:ablation}).

% \vspace{-3mm}
\subsection{Experimental Analysis}
\label{sec:experimental-analysis}
% \vspace{-2mm}
\par\WFclear
\FloatBarrier
\begin{wraptable}{r}{0.65\textwidth}
    \vspace{-12pt}
    \centering
    \scriptsize
    \setlength{\tabcolsep}{2.2pt}
    \renewcommand{\arraystretch}{1.2}
\caption{
% \small
CARE analysis and tool usage in \textit{Recovery}. Percentages in parentheses show case proportions.
}
    \vspace{-3mm}
    \label{tab:care-tool-attribution}
    \begin{tabular}{c@{\hspace{4pt}}lrrr}
        \toprule
        CARE Analysis & Type & Beams (\%) & Hits & Hit Rate \\
        \midrule

        \multirow{7}{*}[2.5pt]{
    \includegraphics[width=0.25\textwidth]
    {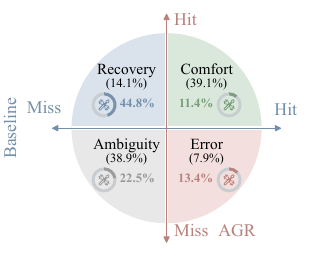}
}
        & \cellcolor{RoyalBlue!9}\textbf{w/ tool call}
        & \cellcolor{RoyalBlue!9}6,304 (44.8\%)
        & \cellcolor{RoyalBlue!9}952
        & \cellcolor{RoyalBlue!9}\textbf{15.1\%} \\

        & \quad GSU
        & 5,664 (40.2\%)
        & 888
        & 15.7\% \\

        & \quad Intent
        & 304 (2.2\%)
        & 38
        & 12.5\% \\

        & \quad Freshness
        & 184 (1.3\%)
        & 15
        & 8.2\% \\

        & \quad Location
        & 152 (1.1\%)
        & 11
        & 7.2\% \\

\cmidrule(l{-2pt}r{0pt}){2-5}

        & \cellcolor{RoyalBlue!9}\textbf{w/o tool call}
        & \cellcolor{RoyalBlue!9}7,776 (55.2\%)
        & \cellcolor{RoyalBlue!9}284
        & \cellcolor{RoyalBlue!9}\textbf{3.7\%} \\

\cmidrule(l{-2pt}r{0pt}){2-5}

        & \textbf{Top-10 beams}
        & 14,080
        & 1,236
        & 8.8\% \\

        \bottomrule
    \end{tabular}

    \vspace{-5pt}
\end{wraptable}

\paragraph{\textit{AGR} recovers baseline failures through selective, precise tool usage.}
To trace the source of \textit{AGR}'s gains,
we partition 10,000 Industry test instances into four quadrants by comparing \textit{AGR} against the tool-free baseline: \textbf{C}omfort (both hit), \textbf{A}mbiguity (both miss), \textbf{R}ecovery (only \textit{AGR} hits), and \textbf{E}rror (only baseline hits). 
As shown in Table~\ref{tab:care-tool-attribution}, 
\textit{Recovery} accounts for 14.1\% of instances versus 7.9\% for \textit{Error}, yielding a net improvement of +6.2pp in Recall@50. 
Notably, tool invocation is heavily concentrated in \textit{Recovery}, where the tool-call rate reaches 44.8\%, far above the 20.6\% overall average.
Within \textit{Recovery}, we further compare tool-invoking and no-tool beams among the most confident top-10 beams for each query: tool-invoking beams achieve a 15.1\% hit rate versus 3.7\% for no-tool beams ($4.1\times$), and every tool type outperforms the no-tool beams.
These findings confirm that \textit{AGR} selectively invokes tools on queries the baseline fails, and that tool-assisted beams are substantially more likely to retrieve the correct item.

\begin{figure}[htbp]
    \centering
    % \vspace{-3mm}
    \begin{minipage}[c]{0.48\linewidth}
        \centering
        \subfigure[
            Comparison between \textit{AGR} and baseline
            on feature-aligned (Fea.) and complementary (Comp.) subsets.
        ]{
            \includegraphics[width=\linewidth]{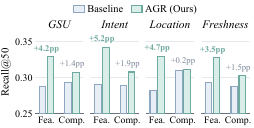}
            \vspace{-3mm}
            \label{fig:tool-subset-gains-macro}
        }
    \end{minipage}%
    \hfill
    % \vspace{-2mm}
    \begin{minipage}[c]{0.48\linewidth}
        \centering
        \vspace{-4mm}
        \subfigure[
\textit{AGR} perplexity relative to Baseline,
        ]{
            \includegraphics[width=\linewidth]{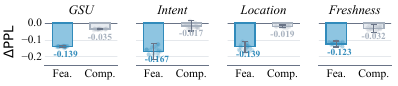}
            \label{fig:when-analysis-agr}
            % \vspace{-5mm}
        }
        \subfigure[
            Comparison of golden-token probability ratio.
        ]{
            \label{fig:where-analysis-agr}
            \begin{minipage}{\linewidth}
            \renewcommand{\arraystretch}{0.92}
                \centering
                \scriptsize
                \setlength{\tabcolsep}{5pt}
                % \vspace{-2mm}
                \resizebox{0.8\linewidth}{!}{%
                \begin{tabular}{lrrrr}
\toprule
\multirow{2}{*}{Feature}
& \multicolumn{2}{c}{Static injection}
& \multicolumn{2}{c}{\textit{AGR}} \\
\cmidrule(lr){2-3}\cmidrule(lr){4-5}
& $r_{i=i^*}$ & $r_{i\neq i^*}$
& $r_{i=i^*}$ & $r_{i\neq i^*}$ \\
\midrule
GSU       & 1.017 & 1.009 & \textbf{1.041 \red{$\uparrow$}} & 1.012 \\
Intent    & 1.032 & 1.040 & \textbf{1.056 \red{$\uparrow$}} & 1.050 \\
Location  & 1.003 & 1.001 & \textbf{1.116 \red{$\uparrow$}} & 1.012 \\
Freshness & 1.008 & 1.010 & \textbf{1.218 \red{$\uparrow$}} & 1.063 \\
\bottomrule
\end{tabular}
                }
            \end{minipage}
        }
    \end{minipage}
    \vspace{-3mm}
    \caption{
Performance analysis of \textit{AGR} under different feature requirements.
    }
    \vspace{-5mm}
    \label{fig:tool-subset-gains}
\end{figure}

\paragraph{\textit{AGR} yields disproportionately larger gains where features matter most.}
To examine whether \textit{AGR}'s improvements are tied to genuine feature relevance, we partition the data into Fea.\ and Comp.\ subsets 
based on whether the target item is characterized by the corresponding feature (Section~\ref{sec:analysis}).
As shown in Figure~\ref{fig:tool-subset-gains-macro}, \textit{AGR} achieves more than twice the recall gain on Fea.\ compared to Comp., confirming that its improvements 
concentrate on feature-dependent queries rather than arising from uniform score shifts.

\paragraph{\textit{AGR} injects evidence at the right moment, avoiding the degradation caused by ill-timed feature injection.}
To examine whether \textit{AGR} resolves the challenge of ``\textbf{when}'' in static injection, we report $\Delta\mathrm{PPL} = \mathrm{PPL}_{\mathrm{AGR}} - \mathrm{PPL}_{\mathrm{w/o\ feature}}$ (analogous to Eq.~\ref{eq:ppl}) on the Fea.\ and Comp.\ set. As shown in Figure~\ref{fig:when-analysis}, static injection lowers PPL on Fea.\ but \emph{raises} it on Comp., exposing its inability to sense when feature injection is actually needed. \textit{AGR}, by contrast, yields negative $\Delta\mathrm{PPL}$ on \emph{both} sets (Figure~\ref{fig:when-analysis-agr}), with a consistently larger reduction on Fea., confirming that it activates tool calls only when evidence is beneficial and withholds them otherwise.

\paragraph{\textit{AGR} concentrates feature signals at the most informative positions rather than spreading them indiscriminately.} 
To examine whether \textit{AGR} resolves the challenge of ``\textbf{where}'' in static injection, we let $i^{*}=\argmax_i\Delta\rho_i$ denote the SID level with the largest change in feature concentration and measure the golden-token probability ratio (Eq.~\ref{eq:prob-gain-ratio}) at $i^{*}$ relative to other levels. As shown in Figure~\ref{fig:where-analysis-agr}, static injection produces nearly the same ratio at $i^{*}$ and elsewhere, whereas \textit{AGR} consistently yields a larger ratio at $i^{*}$ for every feature, with the contrast strongest for Location (1.116 vs.\ 1.012) and Freshness (1.218 vs.\ 1.063), confirming that \textit{AGR} allocates feature influence precisely where it carries the most discriminative power.

\subsection{Online A/B Test}
\label{sec:online-ab-test}

\par\WFclear
\begin{wraptable}{r}{0.5\textwidth}
    \vspace{-12pt}
    \caption{
    % \small 
    Online A/B test results.}
    \label{tab:online-ab-test}
    \vspace{-2mm}
    \centering
    \scriptsize
    \setlength{\tabcolsep}{3pt}
    \renewcommand{\arraystretch}{0.92}
    \begin{tabular}{lll}
        \toprule
        Metric & Change & 95\% CI \\
        \midrule
        Relevance \red{$\uparrow$}              & $+0.205\%$ & $[+0.12\%,\; +0.28\%]$ \\
        Query Reformulation Rate \red{$\downarrow$} & $-0.690\%$ & $[-1.13\%,\;-0.20\%]$ \\
        Search Penetration Rate \red{$\uparrow$}    & $+0.060\%$  & $[+0.03\%,\; +0.14\%]$ \\
        \bottomrule
    \end{tabular}
    \vspace{-8pt}
\end{wraptable}

We deploy \textit{AGR} in our production search system with an average serving latency of 35ms, well within our online budget. We then conduct a live experiment consisting of a 7-day A/A test followed by a 7-day A/B test, with the production baseline and \textit{AGR} each serving 10\% of randomly sampled user traffic. As shown in Table~\ref{tab:online-ab-test}, \textit{AGR} yields consistent gains along three axes of user experience. 
\emph{Relevance}, defined as the fraction of PVs (i.e., search requests) whose top-4 results are strongly relevant to the query, improves by +0.205\%, indicating that \textit{AGR} retrieves more query-relevant items. 
\emph{Query reformulation rate} drops by -0.690\%, showing that users are more often satisfied on the first query and no longer need to rephrase it. Finally, \emph{search penetration rate}, defined as the fraction of users who engage with search, rises by +0.06\%. As an upstream user-behavior metric that \textit{AGR} does not optimize directly, its improvement suggests that better single-query experience propagates into higher willingness to use search.

%% file: sections/main_results_table.tex
\begin{table}[!ht]
% \vspace{-3mm}
    \caption{
        % \small
Overall performance across four datasets.
Best and second-best results are bolded and marked with
\red{$^{*}$} respectively; 
``\textit{Relative gain}'' denotes the relative improvement of \textit{AGR} over TIGER.
Statistical significance is assessed via paired bootstrap tests over
request-level outcomes. 
We construct 1,000 bootstrap resamples with replacement (10,000 requests for Industry from Jun.~23 held-out data; 5,000 for each public benchmark), with the same sampled indices across methods.
% Each Industry resample contains 10,000 requests drawn from the held-out
% data on Jun.~23, while each public-benchmark resample contains 5,000
% requests drawn from the corresponding test set. 
All reported gains of \textit{AGR} are significant at $p<0.05$.
    }
    \label{tab:main-results}
    \centering
    \small
    \setlength{\tabcolsep}{4.2pt}
    \renewcommand{\arraystretch}{0.92}
    \resizebox{\textwidth}{!}{%
    \begin{tabular}{llllllll}
        \toprule
        \multirow{2}{*}{Dataset} & \multirow{2}{*}{Method}
            & \multicolumn{3}{c}{Recall@K} & \multicolumn{3}{c}{MRR@K} \\
        \cmidrule(lr){3-5}\cmidrule(lr){6-8}
        & & Recall@64 & Recall@50 & Recall@32 & MRR@64 & MRR@50 & MRR@32 \\
        \midrule
        \multirow{7}{*}{Industry}
            & TIGER~\citep{TIGER}                  & 0.2957 & 0.2863 & 0.2692 & 0.2257 & 0.2254 & 0.2253 \\
            & TIGER Distill Teacher       & 0.3278 & 0.3169 & 0.2987 & 0.2528 & 0.2524 & 0.2521 \\
            & UniSearch~\citep{Unisearch}                     & 0.3282 & 0.3153 & 0.2979 & 0.2566 & 0.2563 & 0.2559 \\
            & GLASS~\citep{GLASS}                    & 0.3289 & 0.3160 & 0.2986 & 0.2573 & 0.2570 & 0.2566 \\
            & GEMs~\citep{GEMs}                           & 0.3292\red{$^*$} & 0.3165\red{$^*$} & 0.2990\red{$^*$} & 0.2639\red{$^*$} & 0.2635\red{$^*$} & 0.2632\red{$^*$} \\
            & \textbf{\textit{AGR}}   & \textbf{0.3368} & \textbf{0.3264} & \textbf{0.3070} & \textbf{0.2701} & \textbf{0.2699} & \textbf{0.2693} \\
            & \emph{Relative gain over TIGER}   & \cellcolor{cyan!15}+13.90\% & \cellcolor{cyan!15}+14.01\% & \cellcolor{cyan!15}+14.04\% & \cellcolor{cyan!15}+19.67\% & \cellcolor{cyan!15}+19.74\% & \cellcolor{cyan!15}+19.53\% \\
        \midrule
        \multirow{7}{*}{Qilin}
            & TIGER~\citep{TIGER}                  & 0.3330 & 0.3115 & 0.2748 & 0.2923 & 0.2917 & 0.2912 \\
            & TIGER Distill Teacher       & 0.3407 & 0.3198 & 0.2811 & 0.3002 & 0.2999 & 0.2990 \\
            & UniSearch~\citep{Unisearch}                     & 0.3441 & 0.3236 & 0.2838 & 0.3038 & 0.3036 & 0.3025 \\
            & GLASS~\citep{GLASS}                    & 0.3448 & 0.3243 & 0.2845 & 0.3045 & 0.3043 & 0.3032 \\
            & GEMs~\citep{GEMs}                           & 0.3467\red{$^*$} & 0.3267\red{$^*$} & 0.2874\red{$^*$} & 0.3071\red{$^*$} & 0.3068\red{$^*$} & 0.3061\red{$^*$} \\
            & \textbf{\textit{AGR}}   & \textbf{0.3551} & \textbf{0.3324} & \textbf{0.2924} & \textbf{0.3076} & \textbf{0.3069} & \textbf{0.3065} \\
            & \emph{Relative gain over TIGER}   & \cellcolor{cyan!15}+6.64\% & \cellcolor{cyan!15}+6.71\% & \cellcolor{cyan!15}+6.40\% & \cellcolor{cyan!15}+5.23\% & \cellcolor{cyan!15}+5.21\% & \cellcolor{cyan!15}+5.25\% \\
        \midrule
        \multirow{7}{*}{JDSearch}
            & TIGER~\citep{TIGER}                  & 0.5277 & 0.5097 & 0.4749 & 0.3092 & 0.3089 & 0.3080 \\
            & TIGER Distill Teacher       & 0.5360 & 0.5243 & 0.4936 & 0.3292 & 0.3227 & 0.3087 \\
            & UniSearch~\citep{Unisearch}                     & 0.5398 & 0.5315 & 0.5029 & 0.3392 & 0.3295 & 0.3083 \\
            & GLASS~\citep{GLASS}                    & 0.5405 & 0.5322 & 0.5036 & 0.3399\red{$^*$} & 0.3302 & 0.3090 \\
            & GEMs~\citep{GEMs}                           & 0.5663\red{$^*$} & 0.5440\red{$^*$} & 0.5079\red{$^*$} & 0.3338 & 0.3335\red{$^*$} & 0.3326\red{$^*$} \\
            & \textbf{\textit{AGR}}   & \textbf{0.5670} & \textbf{0.5521} & \textbf{0.5163} & \textbf{0.3433} & \textbf{0.3430} & \textbf{0.3422} \\
            & \emph{Relative gain over TIGER}   & \cellcolor{cyan!15}+7.45\% & \cellcolor{cyan!15}+8.32\% & \cellcolor{cyan!15}+8.72\% & \cellcolor{cyan!15}+11.03\% & \cellcolor{cyan!15}+11.04\% & \cellcolor{cyan!15}+11.10\% \\
        \midrule
        \multirow{7}{*}{KuaiSAR}
            & TIGER~\citep{TIGER}                  & 0.4327 & 0.4179 & 0.3953 & 0.3132 & 0.3009 & 0.2990 \\
            & TIGER Distill Teacher       & 0.4448 & 0.4356 & 0.4169 & 0.3332 & 0.3173 & 0.3065 \\
            & UniSearch~\citep{Unisearch}                     & 0.4506 & 0.4444 & 0.4279 & 0.3432 & 0.3255 & 0.3098 \\
            & GLASS~\citep{GLASS}                    & 0.4513 & 0.4451 & 0.4286 & 0.3439 & 0.3262 & 0.3105 \\
            & GEMs~\citep{GEMs}                           & 0.4655\red{$^*$} & 0.4483\red{$^*$} & 0.4293\red{$^*$} & 0.3447\red{$^*$} & 0.3312\red{$^*$} & 0.3288\red{$^*$} \\
            & \textbf{\textit{AGR}}   & \textbf{0.4785} & \textbf{0.4582} & \textbf{0.4335} & \textbf{0.3507} & \textbf{0.3368} & \textbf{0.3348} \\
            & \emph{Relative gain over TIGER}   & \cellcolor{cyan!15}+10.58\% & \cellcolor{cyan!15}+9.64\% & \cellcolor{cyan!15}+9.66\% & \cellcolor{cyan!15}+11.97\% & \cellcolor{cyan!15}+11.93\% & \cellcolor{cyan!15}+11.97\% \\
        \bottomrule
    \end{tabular}
    }
    \vspace{-4mm}
\end{table}

%% file: sections/limitations.tex
\section{Limitations and Future Work}
\label{sec:limitations}
\textit{AGR} interleaves tool invocation with SID generation, but has three limitations. First, its training data are constructed using predefined rules rather than learned policies; future work could apply agentic reinforcement learning to optimize tool use under the sparse rewards common in search and recommendation. Second, the current tool ecosystem is limited and should be extended to more tools and retrieval domains. Third, trajectories are constructed along golden SID prefixes using target-side agreement, creating a gap from free-running inference. Optimizing and evaluating tool decisions under model-generated prefixes is therefore an important future direction.

%% file: sections/conclusion.tex
\section{Conclusion}
\label{sec:conclusion}
We presented \textit{Agentic Generative Retrieval (AGR)}, a tool-augmented paradigm that selectively acquires and incorporates heterogeneous features during hierarchical SID generation. \textit{AGR} determines when, and where features should be introduced based on the request and decoded SID prefix. We train this capability using tool-augmented data constructed from feature applicability and discriminability. Evaluations on an industrial dataset and public benchmarks, together with an online A/B test, demonstrate the effectiveness and efficiency of \textit{AGR}, highlighting tool-augmented generation as a promising direction for large-scale generative retrieval.

%% file: sections/appendix.tex
% \clearpage

\begin{table}[htbp]
    \caption{
    % \small
    Summary of notations.}
    \label{tab:notation}
    \vspace{-3mm}
    \centering
    % \small
    \begin{tabularx}{\linewidth}{@{}lX@{}}
        \toprule
        Notation & Description \\
        \midrule
        \rowcolor{gray!15}
        \multicolumn{2}{l}{\color{black!65}\textbf{Retrieval setting}} \\
        $\ \ \ \mathcal X, x$ & The item corpus and an item in the corpus. \\
        $\ \ \ q,u_{\mathrm{vis}},u$ & The query, model-visible user context and all user context. \\
        % $\ \ \ u$ & all user context available for tool invocation. \\
        $\ \ \ K$ & The number of items retrieved from $\mathcal X$ by beam search. \\
        $\ \ \ \mathrm{Tr}$ & The trie constructed from all legal SIDs in the item corpus. \\
        \rowcolor{gray!15}
        \multicolumn{2}{l}{\color{black!65}\textbf{SID generation and scoring}} \\
        $\ \ \ \mathbf s=[s_1,\ldots,s_L]$ & The length-$L$ SID obtained by quantizing $\mathbf h_x$, where $s_i\in\mathcal S_i$. \\
        $\ \ \ \mathbf s_{\leqslant i}$ & The clean SID prefix $[s_1,\ldots,s_i]$, obtained by removing tool tokens from $\mathbf v_{\mathrm{out}}$. \\
        $\ \ \ \mathbf v_{\mathrm{in}}$ & The decoder input sequence, which may contain an observation token. \\
        $\ \ \ \mathbf v_{\mathrm{out}}$ & The model output sequence, which may contain a tool token. \\
        $\ \ \ \mathbb{P}_\theta$ & Autoregressive conditional probability distribution parameterized by $\theta$. \\
        % $\ \ \ \ell(s_i)$ & Conditional log-probability, i.e., $\log \mathbb{P}_\theta(s_i \mid \mathbf{s}_{<i}, q, u_{\mathrm{vis}})$. \\
        $\ \ \ \pi_\theta(\mathbf s_{\leqslant i})$ & Accumulated log-probability used to score decoded sequences. \\
        $\ \ \ \mathcal{F}$ & The candidate feature set from which subsets are selected for injection. \\
        $\ \ \ \sigma$ & Prefix-aware policy that selects a subset $\mathcal{F}_i \subseteq \mathcal{F}$ at each decoding step. \\
        \rowcolor{gray!15}
        \multicolumn{2}{l}{\color{black!65}\textbf{Tools and observations}} \\
        $\ \ \ \mathcal T,t$ & The set of available tool tokens and a tool identifier. \\
        $\ \ \ \mathcal O,o$ & The observation-token space and an observation token returned by a tool. \\
        $\ \ \ f_t$ & Tool $t$, which returns $o_{t,i}=f_t(q,u,\mathbf s_{\leqslant i})\in\mathcal O$. \\
        $\ \ \ g_t$ & The item-side attribute extractor. \\
        $\ \ \ M$ & Maximum tool-call quota per beam-decoding path. \\
        \rowcolor{gray!15}
        \multicolumn{2}{l}{\color{black!65}\textbf{Interpretive analysis}} \\
        $\ \ \ \rho_i$ & Feature discriminability given prefix $\mathbf{s}_{\leqslant i}$ \\
        \bottomrule
    \end{tabularx}
\end{table}

\section{\textit{AGR} Implementations}
% \addappendixsection{AGR Implementation Details}
\label{app:implementation-details}

This section provides further implementation details of \textit{AGR}.
For datasets with textual features (Industry and Qilin~\citep{chen2025qilin}), we adopt Qwen2.5~\citep{Qwen2.5}
as a parameter-shared dual-encoder model and train it with
\emph{Query-to-Item (Q2I)} supervision. For anonymous datasets
(JDSearch~\citep{liu2023jdsearch} and KuaiSAR~\citep{sun2023kuaisar}),
we instead use TwinBERT~\citep{TwinBERT} to obtain item embeddings. We then
apply hierarchical RQ-KMeans to construct
their Semantic ID (SID) codebooks. Specifically, for the Industry dataset,
which contains billions of recall samples, we use codebook sizes of
$(8192, 4096, 2048, 2048, 2048)$, whereas for the comparatively smaller
public datasets, we use $(512, 256, 128)$. We next describe the
implementation of \textit{AGR} in terms of tools, model architecture,
inference, training-data construction, and training procedure.

\subsection{Datasets and Settings}

Table~\ref{tab:dataset-overview} shows the details of the used datasets, 
Table~\ref{tab:tool-configs} summarizes the two types of tools and their dataset-specific implementations, 
and Table~\ref{tab:key-hyperparameters}
summarizes the training and inference settings.

\begin{table*}[!ht]
    \caption{
    % \small
    Statistics of Industry and three public benchmarks:
    Qilin~\citep{chen2025qilin}, JDSearch~\citep{liu2023jdsearch}, and
    KuaiSAR-L~\citep{sun2023kuaisar}.}
    \label{tab:dataset-overview}
    \vspace{-3mm}
    \centering
    \scriptsize
    \setlength{\tabcolsep}{2.5pt}
    \renewcommand{\arraystretch}{1.00}
    \begin{tabularx}{\linewidth}{@{}
        >{\raggedright\arraybackslash}p{.085\linewidth}
        >{\raggedleft\arraybackslash}p{.052\linewidth}
        >{\raggedleft\arraybackslash}p{.052\linewidth}
        >{\raggedleft\arraybackslash}p{.060\linewidth}
        >{\raggedleft\arraybackslash}p{.066\linewidth}
        >{\raggedright\arraybackslash\hsize=.95\hsize}X
        >{\raggedright\arraybackslash\hsize=1.04\hsize}X
        >{\raggedright\arraybackslash\hsize=1.0\hsize}X@{}}
        \toprule
        & \multicolumn{4}{c}{Retrieval and behavior scale} & & & \\
        \cmidrule(lr){2-5}
        Dataset & Users & Items & Queries & \shortstack{Inter-\\actions}
        & \shortstack[l]{Coverage and\\scenario} & \shortstack[l]{Content and\\feedback} & Use in \textit{AGR} \\
        \midrule
        Industry
        & $\sim$2B & $\sim$1B & $\sim$2.1B$^{\ast}$ & $\sim$2.1B$^{\ast}$
        & Production app search; Jun. 14--23
        & Search logs with impression, click, and engagement supervision
        & Qwen2.5 Q2I embeddings; Jun. 14--20 / 21--22 / 23 for Stage I / Stage II / test \\
        \specialrule{0.25pt}{1.2pt}{1.2pt}
        Qilin
        & 15,482 & 1.98M & 57,188 & 2.50M
        & Xiaohongshu app sessions; search, recommendation, and DQA
        & Text, 5.01M images, and video notes; six engagement labels
        & Qwen2.5 Q2I embeddings; click-history, gender, age, and platform tools \\
        \specialrule{0.25pt}{1.2pt}{1.2pt}
        JDSearch
        & 173,831 & 12.87M & 171,728$^{\dagger}$ & 26.67M
        & JD.com product search; one-year user histories
        & Anonymized queries and product metadata; click, cart, follow, and purchase
        & TwinBERT item embeddings; product-interaction history tool \\
        \specialrule{0.25pt}{1.2pt}{1.2pt}
        KuaiSAR-L
        & 25,877 & 6.89M & 453,667 & 19.66M
        & Kuaishou search + recommendation; 19 days
        & Anonymized queries and item features; exposure, click, play, like, follow, and share
        & TwinBERT item embeddings; short-video viewing-history tool \\
        \bottomrule
    \end{tabularx}
    % \vspace{1pt}

    \parbox{\linewidth}{\scriptsize
    \textbf{\emph{(1) Notes.}} B and M denote billion and million; 
    DQA denotes deep query
    answering; Q2I denotes query-to-item supervision.  $^{\ast}$Industry
    contains approximately 300M query-level interaction records per day, hence
    about 2.1B Stage-I queries/interactions over Jun.~14--20; its user count is
    of the same order of magnitude.  ``Interactions'' denotes logged
    query--item or user--item behavior records (e.g., impressions, clicks, or
    purchases), rather than training actions or tool calls; because event
    definitions differ, these counts are not directly comparable across
    datasets.  $^{\dagger}$JDSearch reports the number of test queries.
    KuaiSAR-L denotes the large release (25,877 users over 19 days), not
    KuaiSAR-small.
\textbf{\emph{(2) Data splits and retrieval corpora.}}
For Industry, we use Jun.~14--20 for Stage-I training,
Jun.~21--22 for Stage-II training, and the held-out data on Jun.~23
for offline evaluation. For statistical testing, we construct 1,000 paired
bootstrap resamples from the Jun.~23 test pool, each containing 10,000
requests sampled with replacement. The same sampled request indices
are used for all compared methods.
For Qilin, we follow its official search-domain split, consisting of
44,024 training instances and 6,192 test instances. Only search-domain
instances are used in our retrieval experiments.
For JDSearch, we follow its original temporal construction, in which
each user's last valid search is used as the request, resulting in
171,728 queries. The preceding interactions are used to construct
the training instances and the product-interaction histories available
to the sequential feature tool. The target interaction and future
behaviors are excluded from the tool state.
Then we pick up 10,000 for testing and use remaining requests for training.
For KuaiSAR-L, we randomly sample 10,000 requests for testing and use
the remaining 443,667 requests for training. For every request, the
viewing-history tool accesses only behaviors preceding that request.
For statistical testing on each public benchmark, we construct 1,000
paired bootstrap resamples, each containing 5,000 test requests sampled
with replacement from the corresponding test set. Identical sampled
request indices are used to evaluate all methods within each resample.
    }

\end{table*}

% \vspace{-5mm}
\begin{table*}[!ht]
    \caption{
    % \small
    Dataset-specific tool configurations. 
    SE denotes \textit{Sequential feature tool}, and ST denotes \textit{Structured feature tool}.}
    \label{tab:tool-configs}
    \vspace{-3mm}
    \centering
    \scriptsize
    \setlength{\tabcolsep}{15pt}
    \renewcommand{\arraystretch}{0.9}
    \resizebox{\linewidth}{!}{%
    \begin{tabular}{llll}
        \toprule
        \textbf{Dataset} & \textbf{Tool} & \textbf{Observations} & \textbf{Type} \\
        \midrule
        \multirow{4}{*}{Industry}
            & Short video viewing behaviors (i.e., GSU) retrieval & Successor SID tokens & SE. \\
            & Query intent classification & 43 intent categories & ST. \\
            & User location lookup & 162 city values & ST. \\
            & Query freshness classification & \{high, medium, low, unknown\} & ST. \\
        \midrule
        \multirow{4}{*}{Qilin}
            & Note clicking behaviors retrieval & Successor SID tokens & SE. \\
            & User gender lookup & \{female, male, unknown\} & ST. \\
            & User age-group lookup & 10 age groups, including unknown & ST. \\
            & Device platform lookup & \{Android, iOS, HarmonyOS\} & ST. \\
        \midrule
        JDSearch & Product interaction behavior & Successor SID tokens & SE. \\
        \midrule
        KuaiSAR & Short video viewing behaviors (i.e., GSU) retrieval & Successor SID tokens & SE. \\
        \bottomrule
    \end{tabular}%
    }
\end{table*}

\begin{table*}[!ht]
    \caption{
    % \small 
    Key training and inference hyperparameters.}
    \vspace{-3mm}
    \label{tab:key-hyperparameters}
    \centering
    \scriptsize
    \setlength{\tabcolsep}{11pt}
    \renewcommand{\arraystretch}{0.95}
    \begin{tabular}{ll@{\hspace{15pt}}ll@{\hspace{15pt}}ll}
        \toprule
        \multicolumn{2}{c}{\textit{Training}}
        & \multicolumn{2}{c}{\textit{AGR Fine-tuning}}
        & \multicolumn{2}{c}{\textit{Inference}} \\
        \midrule

        Optimizer
        & AdamW
        & Voting hit rate $\tau$
        & $0.8$
        & Beam size $K$
        & $\{32,50,64\}$ \\

        Learning rate
        & $1\times10^{-4}$
        & Loss clip $[\alpha_{\min},\alpha_{\max}]$
        & $[0,1]$
        & Tool quota $M$
        & $1$ \\

        Weight decay
        & $0.01$
        & Second stage epochs (Industry / Public)
        & $1 / 10$
        & Top-$E$ (Attn)
        & $5$ \\

        Batch size
        & $1024$
        & Attention heads (Attn)
        & $8$
        & &
        \\

        \bottomrule
    \end{tabular}

    \textbf{\textit{Note}}: Model training and offline evaluation are conducted
    on a V100 GPU cluster, whereas the online A/B test is served on
    RTX 4090 GPUs.
\end{table*}

\subsection{Tool Implementations}
\label{app:tool-implementations}

For structured feature tools, each categorical value is represented by a
learnable observation embedding, initialized by mean-pooling the
pretrained embeddings of the text tokens corresponding to its feature value.
For sequential feature tools, historical GSU items are stored with their
hierarchical SIDs. The Attn variant uses $E=5$ retrieved GSU items and
an 8-head attention module to aggregate their successor-SID representations.
Observation tokens use absolute positional embeddings distinct from those of
autoregressive SID tokens, such that an injected observation conditions the
next prediction without advancing the SID decoding depth.

\iffalse
\label{app:gsu-tool-algorithms}
\begin{algorithm}[H]
\caption{Vote-based prefix-dependent tool call.}
\label{alg:gsu-vote}
\small
\begin{algorithmic}[1]
\Require decoded SID prefix $p=\mathbf s_{\leqslant i}$, historical sequences $\mathcal H$.
\Ensure voted successor observation $o$.
\State $\mathcal M \gets \{x \in \mathcal H : \mathbf s_{x,\leqslant i} = p\}$
\If{$\mathcal M = \emptyset$} \Return \otok{t}{none}
\EndIf
\State $o^* \gets \argmax_{a} |\{x \in \mathcal M : s_{x,i+1} = a\}|$
\State \Return $o^*$
\end{algorithmic}
\end{algorithm}

\begin{algorithm}[H]
\caption{Attn-based sequential feature tool call.}
\label{alg:gsu-indexer}
\small
\begin{algorithmic}[1]
\Require generated SID prefix $p=\mathbf s_{\leqslant i}$, historical sequences $\mathcal H$, encoder \cls\,, query-search top-$E$ size, codebook embeddings $\mathbf{c}$,
Ranking scorer $\phi$.
\Ensure ranked successor observation $o$.
\State $\mathbf z_p \gets \sum_{j=1}^{i} \mathbf{c}(s_j)$, \quad $\mathbf z_x \gets \sum_{j=1}^{i} \mathbf{c}(s_{x,j})$ for $x \in \mathcal{H}$
\Comment{reconstruct vectors}
% \State $\mathcal N \gets $
% \Comment{retrieve top-$E$ similar GSU}
\State $M \gets \{s_{x,i+1} : x \in
\operatorname{Top}E_{x \in \mathcal{H}}\|\mathbf z_p-\mathbf z_x\|
\}$
% \Comment{collect candidate successors}
\If{$M = \emptyset$} \Return \otok{t}{none}
\EndIf
\State $o^* \gets \argmax_{a \in M} \phi(\cls, \mathbf c(s_1), \ldots, \mathbf s_{\leqslant i}, a_1, \ldots, a_E)$
% \Comment{rerank with self-attention}
\State \Return $o^*$
\end{algorithmic}
\end{algorithm}
\fi

\subsection{Model Inference}
% \addappendixsubsection{Model Inference}
\label{app:beam-search}
\paragraph{Inference interface and adaptive interaction.}
We instantiate \textit{AGR} with BART, although the method only requires an
autoregressive interface and also supports decoder-only backbones such as Qwen.
As Figure~\ref{fig:overview} illustrates, its input and output spaces are
asymmetric: the output contains generated SIDs and tool-call decisions, whereas
the decoder input contains SIDs and externally returned observations. A
generated SID is fed back autoregressively; a generated tool token instead
triggers external execution, and the returned observation is fed into the next
decoding step.
This interface represents every tool call and observation with one decoding
position and uses distinct positional encodings to separate observations from
SID decisions. The quota $M$ is maintained separately for every beam-decoding path. A quota of $M$ calls therefore adds at most $M$ decoding steps to each path,
so an $L$-token SID requires at most $L+M+1$ steps including \eos.
% \red{Across the beam set, however, different paths may invoke different tools at different depths; with beam size $K$ and $M=1$, up to $K$ retained paths can each contain one tool call.}
Tool calls are neither fixed to predetermined positions nor required to alternate with SIDs. At each incomplete depth, the model may select either a valid SID successor or an available tool. The method described in Section~\ref{sec:training} determines an admissible position at which to invoke each tool so as to maximize the discriminability of the feature. Theorem~\ref{thm:equivalent-criteria-for-sequential-feature-placement} further characterizes the position at which the feature is most informative w.r.t. sequential features.

% \vspace{-2mm}
\begin{example}[\textit{AGR} Workflow]
\label{ex:agr-workflow}
Figure~\ref{fig:overview} illustrates how \textit{AGR} interleaves SID generation with feature acquisition for the query ``Fighting Zombies''. After the first SID token is generated, the prefix may lead to either the movie or game branch, making the next prediction uncertain. The model therefore emits the intent-tool token \cttok{ToolIntentColor}{I} and receives the game-intent observation $\cotok{ToolIntentColor}{I}{Game}$, which steers decoding toward games while suppressing irrelevant movie candidates.
Further along the game branch, query relevance alone cannot distinguish among candidates such as \emph{Minecraft} and \emph{Plants vs. Zombies}. The model thus emits the GSU-tool token \cttok{ToolGSUColor}{G} to acquire user preferences from historical behaviors. As shown in Figure~\ref{fig:gsu-tool}, \textbf{Vote} retains items whose SIDs exactly match the decoded prefix (A, B, and D) and returns the most frequent successor SID, \stok{3}{7}, indicating a preference for \emph{Minecraft}. In contrast, \textbf{Attn} exploits the residual clustering structure of SIDs: summing the embeddings of the decoded SID tokens reconstructs the prefix representation at the current decoding depth. It then retrieves the Top-$E$ historical behaviors whose item representations are closest to this reconstructed prefix representation, thereby including approximate matches such as C, and aggregates their successor SIDs via lightweight multi-head attention. Both implementations provide fine-grained preference signals that steer decoding toward the user's preferred \emph{Minecraft} branch while suppressing irrelevant candidates.
\end{example}

\vspace{-3mm}
\begin{algorithm}[H]
\caption{Normal trie-based beam search.}
\label{alg:normal-beam-search}
\small
\begin{algorithmic}[1]
\Require query $q$, user features $u$, beam size $K$, SID length $L$, SID trie $\mathrm{Tr}$, generative retrieval model $\mathbb{P}_{\theta}$.
\Ensure top-$K$ relevant items.
\State beams $\mathcal B\gets\{([\bos],\emptyset,0)\}$ \label{line:normal-beam-init}
\For{$j=1$ \textbf{to} $L+1$} \label{line:normal-beam-horizon}
% \Comment{$L$ SIDs, 1 \eos}
    \State next layer candidates $\mathcal C\gets\emptyset$
    \ForAll{$(\mathbf v_{\mathrm{in}},\mathbf v_{\mathrm{out}},z)\in \mathcal{B}$}
        \State $\forall v, \,\ell(v)\gets\log \mathbb{P}_\theta(v\mid q,u_{\mathrm{vis}},\mathbf v_{\mathrm{in}})$, \, \label{line:normal-beam-probs}
            $\mathbf s_{\leqslant i} \gets \mathbf v_{\mathrm{out}} \setminus \{\eos \}$
        \ForAll{$v\in\{\eos\} \cup \, \{\text{next layer SIDs in $\mathrm Tr$ with prefix $\mathbf s_{\leqslant i}$}\}$} \label{line:normal-beam-next}
            \State $\mathcal C\gets\mathcal C\cup\{(\mathbf v_{\mathrm{in}}\oplus v,\mathbf v_{\mathrm{out}}\oplus v,z+\ell(v))\}$ \label{line:normal-beam-add}
            \Comment{$\mathbf v_{\mathrm{out}} \oplus v$: append $v$ to the end of list $\mathbf v_{\mathrm{out}}$}
        \EndFor
    \EndFor
    \State $\mathcal B\gets\operatorname{Top}K_{(\mathbf v_{\mathrm{in}},\mathbf v_{\mathrm{out}},z)\in\mathcal C}(z)$ \label{line:normal-beam-prune}
\EndFor
\State \Return items mapped from $\{\mathbf v_{\mathrm{out}}\setminus\{\eos\}:(\mathbf v_{\mathrm{in}},\mathbf v_{\mathrm{out}},z)\in\mathcal B\}$ \label{line:normal-beam-return}
\end{algorithmic}
\end{algorithm}

\vspace{-4mm}
\begin{algorithm}[H]
\caption{Tool-augmented trie-based beam search.}
\label{alg:tool-beam-search}
\small
\begin{algorithmic}[1]
\Require query $q$, user features $u$, beam size $K$, SID length $L$, SID Trie $\mathrm{Tr}$,
    \color{AlgorithmAccent}
    tool set $\mathcal{T}$, tool quota $M$, \textit{AGR} model $\mathbb P_{\theta}$.
    \color{black}
\Ensure top-$K$ relevant items.
\State Beams $\mathcal B\gets\{([\bos],\emptyset,0)\}$, \, finish set $\mathcal F\gets\emptyset$ \label{line:tool-beam-init}
\color{AlgorithmAccent}
\For{$j=1$ \textbf{to} $L+M+1$} \label{line:tool-beam-horizon}
% \Comment{$L$ SIDs, $M$ tool quota, 1 \eos}
\color{black}
    \State next layer candidates $\mathcal C\gets\emptyset$ \label{line:tool-beam-candidates}
    \ForAll{$(\mathbf v_{\mathrm{in}},\mathbf v_{\mathrm{out}},z)\in\mathcal B\setminus\mathcal F$} \label{line:tool-beam-loop}
        \State $\forall v, \,\ell(v)\gets\log \mathbb{P}_\theta(v\mid q,u_{\mathrm{vis}},\mathbf v_{\mathrm{in}})$ \label{line:tool-beam-probs}
        \ForAll{$v\in\Call{Next}{\mathrm{Tr},\mathbf v_{\mathrm{out}}}$} \label{line:tool-beam-next}
            \color{AlgorithmAccent}
            \If{$v=t\in\mathcal T$} \\
                \quad \quad \quad \quad \quad \quad
                $\mathbf s_{\leqslant i}\gets\mathbf v_{\mathrm{out}}\setminus(\mathcal T\cup\{\eos\})$,
                $o_t\gets f_t(q,u,\mathbf s_{\leqslant i})$,
                $(\mathbf v'_{\mathrm{in}},\mathbf v'_{\mathrm{out}})\gets(\mathbf v_{\mathrm{in}}\oplus o_t,\mathbf v_{\mathrm{out}}\oplus t)$ \label{line:tool-beam-tool}
            \color{black}
            % \vspace{-3mm}
            \Else \,
                $(\mathbf v'_{\mathrm{in}},\mathbf v'_{\mathrm{out}})\gets(\mathbf v_{\mathrm{in}}\oplus v,\mathbf v_{\mathrm{out}}\oplus v)$ \label{line:tool-beam-sid}
                \Comment{$\mathbf v\oplus a$: append $a$ to the end of list $\mathbf v$}
            \EndIf
            \State $\mathcal C\gets\mathcal C\cup\{(\mathbf v'_{\mathrm{in}},\mathbf v'_{\mathrm{out}},z+\ell(v))\}$ \label{line:tool-beam-add}
        \EndFor
    \EndFor
    \State $\mathcal B\gets\operatorname{Top}K_{(\mathbf v_{\mathrm{in}},\mathbf v_{\mathrm{out}},z)\in\mathcal C\cup\mathcal F} (z/|\mathbf v_{\mathrm{out}}|)$ \label{line:tool-beam-prune}
    \State $\mathcal F\gets\{(\mathbf v_{\mathrm{in}},\mathbf v_{\mathrm{out}},z)\in \mathcal B: \mathbf v_{\mathrm{out}}\text{ ends with } \eos\}$ \label{line:tool-beam-finish}
\EndFor
% \color{red}
\State \Return items mapped from $\{\mathbf v_{\mathrm{out}}\setminus(\mathcal T\cup\{\eos\}):(\mathbf v_{\mathrm{in}},\mathbf v_{\mathrm{out}},z)\in\mathcal B\}$ \label{line:tool-beam-return}
% \color{black}
\Function{Next}{$\mathrm{Tr},\mathbf v_{\mathrm{out}}$} \label{line:tool-beam-next-function}
    \State $\mathbf s_{\leqslant i}\gets\mathbf v_{\mathrm{out}}\setminus(\mathcal T\cup\{\eos\})$ \label{line:tool-beam-prefix}
    \If{$i=L$} \, \Return $\{\eos\}$ \label{line:tool-beam-eos} \EndIf
    \State $\mathcal Y\gets \{\text{next-layer SIDs in $\mathrm{Tr}$ with prefix $\mathbf s_{\leqslant i}$}\}$ \label{line:tool-beam-trie}
    \color{AlgorithmAccent}
    \If{$|\mathbf v_{\mathrm{out}}\cap\mathcal T|<M$} \, $\mathcal Y\gets\mathcal Y\cup\mathcal T$ \label{line:tool-beam-budget} \EndIf
    \color{black}
    \State \Return $\mathcal Y$ \label{line:tool-beam-next-return}
\EndFunction
\end{algorithmic}
\end{algorithm}

\paragraph{Beam search.}
The \textbf{\color{AlgorithmAccent}
blue-gray} lines in Algorithm~\ref{alg:tool-beam-search} highlight the core changes relative to the normal trie-based beam search in Algorithm~\ref{alg:normal-beam-search}.
Both algorithms maintain candidates
$(\mathbf v_{\mathrm{in}},\mathbf v_{\mathrm{out}},z)$ and expand trie-valid
SID successors. Their main differences are:
First, conventional decoding uses the fixed $L+1$ horizon
(Algorithm~\ref{alg:normal-beam-search}, line~\ref{line:normal-beam-horizon}),
whereas \textit{AGR} reserves $M$ additional steps and exposes tool tokens only while
quota remains (Algorithm~\ref{alg:tool-beam-search},
line~\ref{line:tool-beam-horizon} and line~\ref{line:tool-beam-budget}). Second, the
conventional update appends every predicted token to both sequences
(Algorithm~\ref{alg:normal-beam-search}, line~\ref{line:normal-beam-add}); \textit{AGR}
instead appends a tool token only to $\mathbf v_{\mathrm{out}}$ and its external
observation only to $\mathbf v_{\mathrm{in}}$, while retaining the conventional
update for SIDs (Algorithm~\ref{alg:tool-beam-search},
lines~\ref{line:tool-beam-tool}). Third, conventional
beam search ranks the current fixed-length expansions by $z$
(Algorithm~\ref{alg:normal-beam-search}, line~\ref{line:normal-beam-prune}); \textit{AGR}
jointly compares active and completed variable-length paths using the
length-normalized score $z/|\mathbf v_{\mathrm{out}}|$
(Algorithm~\ref{alg:tool-beam-search},
lines~\ref{line:tool-beam-prune}). Finally, the
conventional return removes only \eos\ (Algorithm~\ref{alg:normal-beam-search}, line~\ref{line:normal-beam-return}), whereas \textit{AGR} additionally removes tool
tokens to recover the same SID-only output space
(Algorithm~\ref{alg:tool-beam-search}, line~\ref{line:tool-beam-return}).

\subsection{Data Construction and Training}
% \addappendixsubsection{Data Construction and Training}
\label{app:data-construction-training}
\paragraph{Training-data construction.}
Section~\ref{sec:training} presents the overall procedure for constructing tool-augmented supervision. 
% \red{
When multiple tools are simultaneously applicable under a limited tool quota $M$, we prioritize tools with higher coverage in the training data.
Eq.~\ref{eq:structured-feature-concentration} and Eq.~\ref{eq:sequential-feature-concentration} use $\mathcal{X}$ and $\mathcal{D}_{\mathrm{train}}$, respectively, to determine tool insertion positions: the former enables the model, while learning the empirical distribution from $\mathcal{D}_{\mathrm{train}}$, to retrieve the true distribution of the required structural features along SID in $\mathcal{X}$ through tool calls, while the latter intends to inject less popular preferences through target items in $\mathcal{D}_{\mathrm{train}}$.
Example~\ref{ex:training-data-construction} provides a complete worked example illustrating the considerations for \textit{structured feature tools}, including when each tool is applicable and where its observation should be inserted. 

\begin{wrapfigure}[11]{r}{0.30\textwidth}
    \centering
    \vspace{-10mm}
    \includegraphics[width=\linewidth]
        {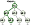}
    \vspace{-6mm}
    \caption{Training data construction example.}
    \vspace{-10mm}
    \label{fig:training-data-example}
\end{wrapfigure}

\begin{example}[Tool-augmented Training Data Construction]
\label{ex:training-data-construction}
Follow the query ``Fighting Zombies'' in Example~\ref{ex:agr-workflow},
Figure~\ref{fig:training-data-example} shows the golden SID path
``$\stok{1}{1}\rightarrow \stok{2}{1}\rightarrow \stok{3}{2} \rightarrow \ldots$'' of the positive item. 
Each path contains 10 items. The numbers of items
with the intent \cotok{ToolIntentColor}{I}{Game} under
\stok{3}{1}, \stok{3}{2}, \stok{3}{3}, and \stok{3}{4} are respectively
$8$, $9$, $2$, and $1$. 
Therefore, along the golden path, the numbers of
\emph{Game} items are $20/40$ under \stok{1}{1}, $17/20$ under
\stok{2}{1}, and $9/10$ under \stok{3}{2}.

\begin{itemize}[leftmargin=*]
    \item \textbf{``When'' is a tool applicable?} 
    Suppose the tool returns
$o_t=\cotok{ToolIntentColor}{I}{Game}$.
It can be invoked because the query intent matches the intent of
the positive item and the intent is absent from the keyword ``Fighting Zombies'' and user features.
    \item \textbf{``Where'' to insert the feature?}
    The concentrations along the golden path are
$\rho_i(o_t)=\log(20/40)$ under \stok{1}{1},
$\rho_{i+1}(o_t)=\log(17/20)$ under \stok{2}{1}, and
$\rho_{i+2}(o_t)=\log(9/10)$ under \stok{3}{2}.
The corresponding gains are $\log(17/20)-\log(20/40)=\log(1.7)$ and $\log(9/10)-\log(17/20)=\log(1.06)$.
Because the first gain is larger, the tool is invoked after \stok{1}{1},
before generating \stok{2}{1}.
\end{itemize}
\end{example}

The example above illustrates how to construct tool-augmented training data. Each tool's insertion depth maximizes the increase of discriminability in Eq.~\ref{eq:observation-position}.
The selected tool token is inserted into the target output, while its observation appears only at the corresponding position in the shifted decoder input.
During inference, each beam path independently decides at which decoding depth to invoke a tool, subject to its own per-path call quota.

\par\WFclear

\section{Interpretive Analysis}
% \addappendixsection{Interpretive Analysis}
\label{app:theoretical-analysis}

\input{sections/theorems}
\input{sections/ablation_study_and_analysis}
\input{sections/related_works}
\input{sections/casestudy}

%% file: sections/theorems.tex
\subsection{An Interpretive View of Dynamic Feature Interaction}
\label{app:dynamic-injection-theory}
% {\color{red}
This section provides an interpretive account of dynamic feature interaction.
It is not intended as a general performance guarantee for \textit{AGR}.
Instead, it explains a simple design intuition: \textbf{as the decoded SID prefix
becomes more fine-grained, the information that is useful for predicting the next SID token may also change.}

\begin{theorem}[Contextual Incremental Information]
    A feature observation $o_t$ is useful w.r.t. model $\mathbb P_{\theta}$ only if it adds information beyond the current
context (i.e., $q, u_{\mathrm{vis}}$ and $\mathbf{s}_{\leqslant i}$).
\label{thm:contextual-incremental-information}
\end{theorem}

\begin{proof}
Consider the conditional mutual information at the current context by definition:
\begin{align}
&I_{\theta}(s_{i+1}; o_t \mid q, u_{\mathrm{vis}}, \mathbf{s}_{\leqslant i}) \nonumber\\
&\quad =
\mathbb E
\left[
\mathbb D_{\mathrm{KL}}\!\left(
\mathbb P_\theta(s_{i+1}\mid q, u_{\mathrm{vis}}, \mathbf{s}_{\leqslant i}, o_t)
\,\middle\|\,
\mathbb P_\theta(s_{i+1}\mid q, u_{\mathrm{vis}}, \mathbf{s}_{\leqslant i})
\right)
\right].
\label{eq:feature-cmi-kl}
\end{align}
If this quantity is zero, the nonnegativity of the KL divergence implies that
\begin{equation}
\mathbb P_\theta(s_{i+1}\mid q, u_{\mathrm{vis}}, \mathbf{s}_{\leqslant i}, o_t)
=
\mathbb P_\theta(s_{i+1}\mid q, u_{\mathrm{vis}}, \mathbf{s}_{\leqslant i})
\quad \text{almost surely}.
\label{eq:feature-no-effect}
\end{equation}
Hence, $o_t$ provides no additional predictive information for the next SID token at the current context (i.e., $q, u_{\mathrm{vis}}, \mathbf{s}_{\leqslant i}$). The contrapositive proves the theorem.
\end{proof}

Although conditional mutual information is nonnegative in expectation, the predictive gain of a realized observation can vary across contexts. This consideration is particularly important in industrial search and recommendation systems, where numerous features are available to characterize complex and evolving user preferences, yet generative retrieval models can incorporate only limited additional information during decoding because of efficiency and scalability constraints. Feature injection must therefore be selective, retaining observations that improve prediction at the current prefix while avoiding those that provide no gain or reduce the probability of the correct next SID token. Our method implements this context-dependent selection through tool invocation and observation injection (Section~\ref{sec:method}); next, we analyze this mechanism through experiments on tool selection and invocation.

\paragraph{Generative Retrieval Knapsack Problem (GR-KP).}
In \textit{AGR}, the feature selection is implemented through tool interaction. 
Given the current context, predicting a tool token \(t\in\mathcal T\) invokes its associated function \(f_t\) and returns an observation
\(o_t=f_t(q,u,\mathbf{s}_{\leqslant i})\). The returned observation token is then injected into
the decoder to influence subsequence prediction.
We formulate this feature injection as a \textit{Generative Retrieval Knapsack Problem (GR-KP)}. At context \(q, u_{\mathrm{vis}}\) and $\mathbf{s}_{\leqslant i}$, invoking tool \(t\) incurs cost \(c_t>0\) and yields context-dependent value \(v_t(\mathbf{s}_{\leqslant i})\), which summarizes
the contribution of its returned observation to decoding process. Let
\(\gamma_t\in[0,1]\) denote the allocation assigned to each tool \(t\). Under a
finite model capacity \(C_\theta\), the feasible allocation set $\Gamma$ and its
value $V$ are
\begin{equation}
% \small
\Gamma =
\left\{
\boldsymbol{\gamma} = (\gamma_t)_{t\in\mathcal T}
\,\middle|\,
0 \leqslant \gamma_t \leqslant 1, \
% \ \forall t\in\mathcal T, \ 
\sum_{t\in\mathcal T} c_t \gamma_t \leqslant C_\theta
\right\},
\quad
V(\mathbf{s}_{\leqslant i},\boldsymbol{\gamma})
=
\sum_{t\in\mathcal T} \gamma_t v_t(\mathbf{s}_{\leqslant i}).
\label{eq:grkp-definition}
\end{equation}

Here, fractional allocation provides a compact estimator of the benefit of feature selection in a capacity-limited generative model. As discussed in Section~\ref{sec:analysis}, \emph{static injection} assigns a single allocation \(\boldsymbol{\gamma}\) shared across all prefixes, whereas \emph{dynamic injection} assigns a separate allocation \(\boldsymbol{\gamma}_i\) to the corresponding prefix \(\mathbf{s}_{\leqslant i}\).
Thus, the optimal expected cumulative utility under \textit{static} and \textit{dynamic injection} are
\begin{equation}
% \small
\mathcal{J}_{\mathrm{static}}
=
\max_{\boldsymbol{\gamma}\in\Gamma}
\mathbb{E}\left[\sum_i V(\mathbf{s}_{\leqslant i},\boldsymbol{\gamma})\right],
\qquad
\mathcal{J}_{\mathrm{dynamic}}
=
\mathbb{E}\left[
\sum_i\max_{\boldsymbol{\gamma}_i\in\Gamma}
V(\mathbf{s}_{\leqslant i},\boldsymbol{\gamma}_i)
\right].
\label{eq:allocation-objectives}
\end{equation}

\begin{theorem}[Dynamic Injection is Better]
\label{thm:dynamic-allocation-is-better}
When the usefulness of tool-derived features depends on the prefix
\(\mathbf{s}_{\leqslant i}\), dynamic injection is at least as effective as
static injection, i.e.,
\(
\mathcal{J}_{\mathrm{dynamic}}
\geq
\mathcal{J}_{\mathrm{static}}.
\)
\end{theorem}

\begin{proof}
Recall that static injection selects one allocation
\(\boldsymbol{\gamma}\in\Gamma\) and uses it for every prefix:
$
\mathcal{J}_{\mathrm{static}}
=
\max_{\boldsymbol{\gamma}\in\Gamma}
\mathbb{E}\!\left[
    \sum_i V(\mathbf{s}_{\leqslant i},\boldsymbol{\gamma})
\right].
$
A dynamic injection is allowed to choose an allocation
\(\boldsymbol{\gamma}_i(\mathbf{s}_{\leqslant i})\in\Gamma\) after observing the
current prefix. In particular, it can imitate any static policy by setting
$
\boldsymbol{\gamma}_i(\mathbf{s}_{\leqslant i})
=
\boldsymbol{\gamma},
\text{for all } \mathbf{s}_{\leqslant i}.
$
Thus, every static allocation is feasible for the dynamic injection, and therefore
\begin{equation}
    % \small
\mathcal{J}_{\mathrm{dynamic}}
=
\mathbb{E}\!\left[
    \sum_i
    \max_{\boldsymbol{\gamma}_i\in\Gamma}
    V(\mathbf{s}_{\leqslant i},\boldsymbol{\gamma}_i)
\right]
\geq
\max_{\gamma \in \Gamma}
\mathbb{E}\!\left[
    \sum_i
    V(\mathbf{s}_{\leqslant i},\boldsymbol{\gamma})
\right]
=
\mathcal{J}_{\mathrm{static}}.
\end{equation}
\end{proof}

This indicates that \textit{dynamic injection} is a more dynamic and superior feature selection planning compared to \textit{static injection}.

\begin{theorem}[Equivalent Criteria for Sequential Feature Placement]
\label{thm:equivalent-criteria-for-sequential-feature-placement}
For a sequential feature, the position with the largest probability gain
for the target next SID is equivalently the position that maximizes
\(
\rho_{i+1}(o_{t,i})-\rho_i(o_{t,i})
\)
and minimizes
\(
\rho_i(o_{t,i})
\).
\end{theorem}

\begin{proof}
Assume that the distribution learned by a sufficiently trained model
approximates the empirical distribution of the training data:
$
\mathbb{P}_{\theta}
\sim
\mathbb{P}_{\mathcal{D}_{\mathrm{train}}}.
$
Following Eq.~\ref{eq:prob-gain-ratio}, consider the logarithm of the
probability-gain ratio \(r_i\) for predicting \(s_{i+1}^+\) after
conditioning on the observation \(o_{t,i}\) in addition to the prefix
\(\mathbf{s}_{\leqslant i}^+\):
\begin{align}
\log r_i
&=
\log \left[
\frac{
\mathbb{P}_{\theta}
\left(
s_{i+1}^{+}
\mid
\mathbf{s}_{\leqslant i}^{+},o_{t,i}
\right)
}{
\mathbb{P}_{\theta}
\left(
s_{i+1}^{+}
\mid
\mathbf{s}_{\leqslant i}^{+}
\right)
}
\right]
\approx
\log \left[
\frac{
\mathbb{P}_{\mathcal{D}_{\mathrm{train}}}
\left(
s_{i+1}^{+}
\mid
\mathbf{s}_{\leqslant i}^{+},o_{t,i}
\right)
}{
\mathbb{P}_{\mathcal{D}_{\mathrm{train}}}
\left(
s_{i+1}^{+}
\mid
\mathbf{s}_{\leqslant i}^{+}
\right)
}
\right]
.
\label{eq:sequential-log-gain}
\end{align}

By Bayes' rule, we have
\begin{align}
\mathbb{P}_{\mathcal{D}_{\mathrm{train}}}
\left(
s_{i+1}^{+}
\mid
\mathbf{s}_{\leqslant i}^{+},o_{t,i}
\right)
=
\frac{
\mathbb{P}_{\mathcal{D}_{\mathrm{train}}}
\left(
o_{t,i}
\mid
\mathbf{s}_{\leqslant i}^{+},s_{i+1}^{+}
\right)
\cdot
\mathbb{P}_{\mathcal{D}_{\mathrm{train}}}
\left(
s_{i+1}^{+}
\mid
\mathbf{s}_{\leqslant i}^{+}
\right)
}{
\mathbb{P}_{\mathcal{D}_{\mathrm{train}}}
\left(
o_{t,i}
\mid
\mathbf{s}_{\leqslant i}^{+}
\right)
}.
\label{eq:sequential-bayes}
\end{align}

Substituting Eq.~\ref{eq:sequential-bayes} into
Eq.~\ref{eq:sequential-log-gain} and canceling the common term yields
\begin{align}
\log r_i
&\approx
\log
\mathbb{P}_{\mathcal{D}_{\mathrm{train}}}
\left(
o_{t,i}
\mid
\mathbf{s}_{\leqslant i}^{+},s_{i+1}^{+}
\right)
-
\log
\mathbb{P}_{\mathcal{D}_{\mathrm{train}}}
\left(
o_{t,i}
\mid
\mathbf{s}_{\leqslant i}^{+}
\right)
\nonumber\\
&=
\log
\mathbb{P}_{\mathcal{D}_{\mathrm{train}}}
\left(
o_{t,i}
\mid
\mathbf{s}_{\leqslant i+1}^{+}
\right)
-
\log
\mathbb{P}_{\mathcal{D}_{\mathrm{train}}}
\left(
o_{t,i}
\mid
\mathbf{s}_{\leqslant i}^{+}
\right)
\nonumber\\
&=
\rho_{i+1}(o_{t,i})-\rho_i(o_{t,i}).
\label{eq:sequential-rho-gain}
\end{align}

Since the logarithm is strictly increasing, and the distributional
approximation preserves the ordering across admissible positions,
\begin{equation}
\operatorname*{arg\,max}_{i} r_i
\;\Longleftrightarrow\;
\operatorname*{arg\,max}_{i}
\left[
\rho_{i+1}(o_{t,i})-\rho_i(o_{t,i})
\right].
\end{equation}

Moreover, for every correct sequential prediction satisfying
Eq.~\ref{eq:Pseq}, having $o_{t,i}=s_{i+1}^{+}.$
Under the next-token prediction setting, given the prefix
\(\mathbf{s}_{\leqslant i+1}^{+}\),
\(\rho_{i+1}(o_{t,i})\) remains unchanged and can therefore be
treated as a constant \(C'\),
indicating that
\begin{equation}
\rho_{i+1}(o_{t,i})-\rho_i(o_{t,i})
=
C'-\rho_i(o_{t,i}),
\end{equation}
and hence
\begin{equation}
\operatorname*{arg\,max}_{i}
\left[
\rho_{i+1}(o_{t,i})-\rho_i(o_{t,i})
\right]
\;\Longleftrightarrow\;
\operatorname*{arg\,min}_{i}
\rho_i(o_{t,i}).
\end{equation}

Therefore,
\begin{equation}
\operatorname*{arg\,max}_{i} r_i
\;\Longleftrightarrow\;
\operatorname*{arg\,max}_{i}
\left[
\rho_{i+1}(o_{t,i})-\rho_i(o_{t,i})
\right]
\;\Longleftrightarrow\;
\operatorname*{arg\,min}_{i}
\rho_i(o_{t,i})
.
\end{equation}
\end{proof}

\begin{tcolorbox}[
    colback=green!5!white,
    colframe=green!5!white,
    boxrule=0pt,
    arc=0pt,
    left=6pt,
    right=6pt,
    top=5pt,
    bottom=5pt
]
\textbf{Takeaway.}
Theorem~\ref{thm:contextual-incremental-information} shows that a tool is useful only when it provides information beyond
the current prefix. Theorem~\ref{thm:dynamic-allocation-is-better} shows that adapting tool selection to each prefix
is at least as effective as using a fixed selection. Theorem~\ref{thm:equivalent-criteria-for-sequential-feature-placement} shows that, for
sequential features, the position with the largest probability gain can be
identified by maximizing the change in \(\rho\), equivalently by minimizing
its current value. Together, these results motivate when, which, and where
tools should be invoked during decoding.
\end{tcolorbox}

%% file: sections/ablation_study_and_analysis.tex
% \vspace{-3mm}
\clearpage
\section{Further Ablation study and Analysis}
\label{app:ablation}
\begin{wraptable}{r}{0.46\linewidth}
    \centering
    \vspace{-4mm}
    \caption{
    % \small 
    Observation analysis.}
    \label{tab:rq3-observation-analysis}
    \scriptsize
    \setlength{\tabcolsep}{10pt}
    \renewcommand{\arraystretch}{0.92}
    \begin{tabular}{lc}
        \toprule
        Method & Recall\@50 \\
        \midrule
        Baseline
        & 0.2863 \\
        \textit{AGR} w/ shared observation
        & 0.2901 \\
        \textit{AGR}
        & \textbf{0.3264} \\
        \bottomrule
    \end{tabular}
    
    \textit{\textbf{Note:}} All tools are enabled in this ablation.

    \vspace{-3mm}
\end{wraptable}

\paragraph{Observations contribute genuine feature information, not merely extra decoding steps.} To verify that performance gains stem from the specific feature information carried by tool observations rather than from additional reasoning tokens, we replace per-observation tokens with a single shared learnable token, equivalent to a late-reasoning approach (\textit{AGR} w/ shared observations). As shown in Table~\ref{tab:rq3-observation-analysis}, this reasoning variant already outperforms the Baseline, confirming that extra decoding steps alone provide a non-trivial benefit. Yet \textit{AGR} with distinct observation tokens achieves further gains, demonstrating that the observation tokens carry substantive feature information beyond what additional steps can provide. 

\begin{wraptable}{r}{0.45\textwidth}
    \vspace{-10pt}
    \caption{
    % \small 
    Structured injection vs. \textit{AGR}.}
    \label{tab:apprq5-structured-encoding}
    \centering
    \scriptsize
    \setlength{\tabcolsep}{7pt}
    \renewcommand{\arraystretch}{0.90}
    \begin{tabular}{lc}
        \toprule
        Method & Recall@50 \\
        \midrule
        Baseline & 0.2863 \\
        Static encoding for all structured features & 0.3013 \\
        \textit{AGR} w/ all structured feature tools only & \textbf{0.3137} \\
        \bottomrule
    \end{tabular}
    \vspace{-6pt}
\end{wraptable}
\paragraph{Structured feature tools outperform the direct encoding approach.} 
In the \textit{AGR} model, structured feature tools are incorporated into the decoding process through a token, depending on whether the relevant information is required at the current decoding step. An intuitive alternative is to directly integrate this information into the model's encoding to evaluate its potential benefits. As shown in Table~\ref{tab:apprq5-structured-encoding}, directly encoding the relevant information improves performance compared to the baseline (from 0.2863 to 0.3013). However, the encoding strategy employed by \textit{AGR} achieves even better results, with a score of 0.3137. This demonstrates that introducing relevant information on demand during the decoding process reduces the inclusion of noisy information, thereby enabling more efficient information encoding.

\par\WFclear
\begin{wraptable}{r}{0.45\textwidth}
    \vspace{-12pt}
    \caption{
    % \small
    Effect of backbone and model scaling on generative \textit{AGR}.
    }
    \label{tab:backbone-scaling}
    \centering
    \scriptsize
    \setlength{\tabcolsep}{12pt}
    \renewcommand{\arraystretch}{0.92}
    \begin{tabular}{lrr}
        \toprule
        Method & Recall@50 & MRR@50 \\
        \midrule

        \rowcolor{black!10}
        \multicolumn{3}{l}{
            \color{black!65}\textbf{\textit{BART}}
        } \\
        UniSearch & 0.3153 & 0.2563 \\
        \textbf{\textit{AGR}}       & \textbf{0.3264} & \textbf{0.2699} \\
        \midrule
        \addlinespace[2pt]
        \rowcolor{black!10}
        \multicolumn{3}{l}{
            \color{black!65}\textbf{\textit{Qwen3-0.6B}}
        } \\
        UniSearch & 0.3250 & 0.2772 \\
        \textbf{\textit{AGR}}       & \textbf{0.3347} & \textbf{0.2784} \\
        \midrule
        \addlinespace[2pt]
        \rowcolor{black!10}
        \multicolumn{3}{l}{
            \color{black!65}\textbf{\textit{Qwen3-1.7B}}
        } \\
        UniSearch & 0.3288 & 0.2843 \\
        \textbf{\textit{AGR}}       & \textbf{0.3350} & \textbf{0.2876} \\
        \midrule
        \addlinespace[2pt]
        \rowcolor{black!10}
        \multicolumn{3}{l}{
            \color{black!65}\textbf{\textit{Qwen3-4B}}
        } \\
        UniSearch & 0.3369 & 0.2960 \\
        \textbf{\textit{AGR}}       & \textbf{0.3444} & \textbf{0.2989} \\

        \bottomrule
    \end{tabular}
    \vspace{-6pt}
\end{wraptable}

\paragraph{\textit{AGR}'s gains persist across backbone architectures and model scales.} To examine the effects of backbone architecture and model scale, we evaluate UniSearch and \textit{AGR} using BART~\citep{BART} and Qwen3 backbones with 0.6B, 1.7B, and 4B parameters~\citep{Qwen3}. As shown in Table~\ref{tab:backbone-scaling}, Qwen3-0.6B outperforms the similarly sized BART backbone, suggesting that stronger pretrained language modeling provides benefits beyond parameter count alone. Scaling Qwen3 generally improves performance, with the 4B backbone achieving the best overall results. More importantly, \textit{AGR} consistently outperforms UniSearch under every matched backbone, demonstrating that its improvements cannot be attributed solely to model capacity. These results confirm that backbone scaling and \textit{AGR} provide complementary gains: larger pretrained models offer a stronger foundation, while explicit feature selection and targeted injection further improve SID generation across architectures and scales.

\par\WFclear
\FloatBarrier
\begin{wraptable}{r}{0.45\textwidth}
    \vspace{-12pt}
    \caption{
    % \small 
    GSU observation fusion strategy.}
    \label{tab:fusion-strategy}
    \centering
    \scriptsize
    % \small
    \renewcommand{\arraystretch}{0.92}
        \centering
        % \textit{Observation fusion}\\[1pt]
        \setlength{\tabcolsep}{12pt}
        \begin{tabular}{lc}
            \toprule
            Observation fusion variant & Recall@50 \\
            \midrule
            \emph{Mean}  & 0.3167 \\
            \emph{Vote} & 0.3237 \\
            \emph{Attn} & 0.3313 \\
            \bottomrule
        \end{tabular}
        
    \textbf{\textit{Note}}:
    only GSU tool is enable for this ablation.
    % \end{minipage}
    \vspace{-8pt}
\end{wraptable}

\paragraph{The impact of fusion modes in sequential feature tools: attention-based fusion is most accurate; voting is the practical default.} 
As discussed in Section~\ref{sec:agr-method}, we implement two fusion modes for sequential feature tools (\emph{Vote} and \emph{Attn}) and introduce \emph{Mean} as an additional baseline. \emph{Mean} filters GSUs by prefix matching and averages their embeddings uniformly, discarding candidate importance variations; as expected, it yields the weakest results (Table~\ref{tab:fusion-strategy}). \emph{Attn} achieves the best performance by relaxing strict prefix matching and learning adaptive fusion weights, confirming that candidate importance is both heterogeneous and learnable. Nevertheless, since \emph{Vote} reaches comparable performance at substantially lower deployment cost, we adopt it as the default throughout this paper.

\par\WFclear
\begin{wrapfigure}{r}{0.45\linewidth}
    \vspace{-8pt}
    \centering
    \subfigure[Tool-loss rescaling.]{
        \includegraphics[width=0.56\linewidth]{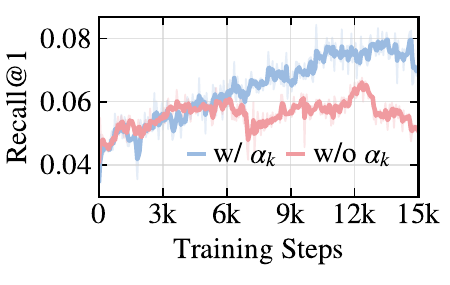}
        \label{fig:tool-loss-rescaling}
    }%
    \hspace{-2mm}%
    \subfigure[Loss scale.]{
        \includegraphics[width=0.40\linewidth]{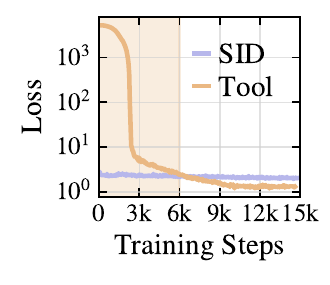}
        \label{fig:loss-scale}
    }
    \vspace{-10pt}
    \caption{Tool-loss rescaling and loss scale.}
    \label{fig:training-dynamics}
    \vspace{-10pt}
\end{wrapfigure}

\paragraph{Tool loss rescaling prevents newly introduced tool tokens from disrupting SID semantics.}
Stage-II training introduces tool tokens on top of the SID generation ability learned in Stage-I, creating a scale mismatch between the pretrained SID objective and the newly added tool objective. To test whether this mismatch matters, we compare training with and without the batch-wise rescaling coefficient $\alpha_k$ in Eq.~\ref{eq:agr_loss}. As shown in Figure~\ref{fig:tool-loss-rescaling}, rescaling consistently improves Recall@1, while Figure~\ref{fig:loss-scale} shows a clear early-stage gap between SID- and tool-token loss magnitudes. This confirms that optimizing tool tokens at their raw scale can disturb the pretrained SID semantics, whereas $\alpha_k$ calibrates the two objectives and enables stable tool-enhanced semantic learning.

%% file: sections/related_works.tex
\vspace{-3mm}
\section{Related Works}
\label{app:related-works}
\vspace{-2mm}
\paragraph{Generative retrieval and semantic identifiers.}
Distinct from traditional discriminative retrieval~\citep{xie2026crops,zheng2024full}, generative retrieval reformulates retrieval as conditional sequence generation over discrete identifiers~\citep{Unisearch,tay2022transformermemorydifferentiablesearch}. Early work focused on document retrieval, where DSI stores query--document associations in model parameters~\citep{tay2022transformermemorydifferentiablesearch}, NCI introduces structured identifiers and prefix-aware decoding~\citep{wang2023neuralcorpusindexerdocument}, and SEAL uses document-derived textual access paths~\citep{bevilacqua2022autoregressivesearchenginesgenerating}. These studies established that the identifier space is learnable and that errors at different decoding positions have different retrieval consequences~\citep{sun2023learning,zeng2023scalableeffectivegenerativeinformation}. Generative recommendation extends this idea to multi-token item semantic identifiers (SIDs). TIGER quantizes item representations with RQ-VAE and generates hierarchical SIDs~\citep{TIGER}, making coarse-to-fine constrained decoding a standard formulation. Subsequent work shows that SID quality depends on semantic consistency, discrimination, codebook utilization, and tree balance~\citep{SemanticSnapchat}, while autoregressive SID trees may also restrict ranking expressiveness~\citep{SubtleQuery}. \textit{AGR} builds on this SID-based generative retrieval paradigm, but shifts the focus from how identifiers are constructed to how external evidence should interact with SID generation.
\vspace{-2mm}
\paragraph{Feature conditioning in generative retrieval.}
Real-world GR systems rely on heterogeneous auxiliary information, including user profiles, behavioral histories, scenario context, freshness signals, and intent features. Existing methods usually incorporate such information before decoding or through a fixed injection pattern~\citep{wang2025lad}. DiffusionGS performs query-conditioned denoising over candidate items~\citep{FusionGS}; DualGR separates long- and short-term user interests~\citep{yi2026dualgrgenerativeretrievallong}; GLASS retrieves user histories conditioned on coarse SIDs~\citep{GLASS}; and OneReason partitions behavioral histories into temporal streams~\citep{Onereason}. These methods show that request-aware evidence selection is important, and they move beyond globally encoding all context indiscriminately. However, their injection mechanisms are still largely predetermined: evidence is selected before generation, kept fixed after a coarse decision, or injected at predefined network locations. They do not formulate feature acquisition itself as an explicit decoding-time action. \textit{AGR} differs by expanding the generative action space with tool tokens, allowing the model to decide whether a feature is needed, which feature to acquire, and where in the SID hierarchy the acquired evidence should affect subsequent generation.
\vspace{-2mm}
\paragraph{Industrial GR systems and optimization.}
Industrial systems pursue GR for unification, cross-domain reuse, and serving efficiency. UniSearch jointly trains generators and encoders for video and live search~\citep{Unisearch}; OneSearch develops task-aware SID construction for e-commerce~\citep{Onesearch}; IntSR shares models across search and recommendation~\citep{yan2025intsr}; OneRec unifies retrieval and ranking~\citep{Onerec}; and OneMall extends GR across multiple scenarios~\citep{zhang2026onemallarchitecturescenarios}. Complementary work improves SID construction, collision control, and long-term maintenance~\citep{wang2026intrrframeworkintegratingsid,fu2026forgeformingsemanticidentifiers,baikalov2026mitigatingcollaborativesemanticid,fang2026generative}, and generative interfaces have also been extended to reranking and query recommendation~\citep{yang2026breakinglikelihoodtrapconsistent,shao2025greatguidingquerygeneration}. Beyond maximum-likelihood training, preference-oriented methods align GR with ranking quality or business value: UniSearch uses ranking signals~\citep{Unisearch}, while related methods apply preference optimization to e-commerce~\citep{li2024generativeretrievalpreferenceoptimization}, adapt DPO to hierarchical SIDs~\citep{chen2026raddporobustadaptivedenoising2}, use listwise optimization~\citep{fu2026rankgrrankenhancedgenerativeretrieval}, generative flow networks~\citep{wang2026gflowgrfinetuninggenerativerecommendation}, token-level signals~\citep{lin2026tokenlevelcollaborativealignmentllmbased}, iterative preference construction~\citep{Onerec}, page-wise supervision~\citep{zou2026genrecpreferenceorientedgenerativeframework}, and list-level value optimization~\citep{xue2026generativerecommendationlargescaleadvertising}. Efficiency methods accelerate decoding through draft verification~\citep{wang2026nezhazerosacrificehyperspeeddecoding}, process reward models~\citep{guo2026promiseprocessrewardmodels}, and load-aware computation~\citep{xue2026generativerecommendationlargescaleadvertising}. These efforts optimize different components of large-scale GR pipelines, whereas \textit{AGR} targets a complementary bottleneck: avoiding noisy and expensive static feature injection by acquiring useful features on demand during decoding.
\vspace{-2mm}
\paragraph{Reasoning and agentic retrieval.}
Recent work introduces intermediate reasoning to improve intent understanding in search and recommendation. OneSearch-V2 constructs keyword-based chains of thought~\citep{OneSearch-V2}; OneRec-Think interleaves textual reasoning with generation~\citep{Onerec-think}; SCoTER transfers structured reasoning from LLMs~\citep{jiang2026scoterstructuredchainofthoughttransfer}; DiffuReason combines latent thinking with diffusion refinement~\citep{jiang2026diffureasonbridginglatentreasoning}; and OneReason organizes recommendation around perception and cognition~\citep{Onereason}. Agentic recommendation further allows models to actively acquire external evidence. KP-Agent combines LLM agents with domain tools~\citep{long2025kpagentkeywordpruningsponsored}, and ChainRec learns to route and terminate tool chains~\citep{li2026chainrecagenticrecommenderlearning}. However, general-purpose LLM planning typically introduces sequential inference, external orchestration, and tool-chain latency, which are difficult to accommodate in latency-constrained retrieval. Unlike these agentic recommenders, \textit{AGR} treats feature access as a lightweight action inside native constrained SID decoding. Tool invocation is not a separate planning loop, but an interleaved part of beam search that directly reshapes the probability distribution over subsequent SID tokens.

%% file: sections/casestudy.tex
% \clearpage
% \vspace{-3mm}
\section{Case Study}
\label{sec:case-study}

\definecolor{CaseStudyFill}{HTML}{F5F8FC}
\definecolor{CaseStudyBorder}{HTML}{6E8197}
\definecolor{CaseFreshnessColor}{HTML}{FFAA00}
\definecolor{CaseIntentColor}{HTML}{FFADAD}

% \vspace{-3mm}
\begin{tcolorbox}[
    breakable,
    colback=CaseStudyFill,
    colframe=CaseStudyBorder,
    boxrule=0.7pt,
    arc=2pt,
    left=7pt,right=7pt,top=7pt,bottom=7pt,
    before skip=7pt,after skip=7pt,
    fontupper=\small]

\paragraph{Case overview.}
We present six representative requests to illustrate how \textit{AGR}
selectively invokes user- and query-dependent tools during generation:
\emph{World Cup Portugal} combines GSU, Freshness, and Intent;
\emph{Typhoon News} mainly uses Freshness; \emph{Python Course} and
\emph{Study Application} use GSU to infer the user's learning and application
goals; \emph{Jeju Island} mainly uses Intent; and \emph{Fireworks Festival}
uses GSU to capture visual and aesthetic preferences. We first detail the
tool-guided inference of \emph{World Cup Portugal}; the other five cases are
shown afterward.

\vspace{+5mm}
\makeatletter\def\@captype{figure}\makeatother
\centering
\includegraphics[width=\linewidth]{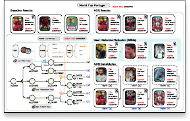}
\vspace{-9mm}
\caption{Case 1: Top-3 results for ``World Cup Portugal''}
\label{fig:world-cup-portugal}

% \par\medskip
% \raggedright
\begin{example}[Detailed inference for ``World Cup Portugal'']
The user had recently consumed several football videos and then issued the
request ``World Cup Portugal''.  The historical behaviors suggest an interest not only in
the broad topic, but specifically in recent Portugal matches and Cristiano
Ronaldo.  The baseline captures the coarse topic, yet mixes a recent
Ronaldo video with generic highlights and an older Portugal--DR Congo match (Figure~\ref{fig:world-cup-portugal}).
It therefore misses the request's personalized, temporal, and intent-level
requirements.
\textit{AGR} resolves these requirements through three tool-guided decoding
paths. The showed calls below occur independently on three different beam paths:
\begin{itemize}[leftmargin=*,itemsep=3pt,topsep=3pt]
    \item \textcolor[HTML]{70C0C7}{\textbf{Personalization through GSU.}}
    At the prefix \([\stok{1}{6421},\stok{2}{11037}]\), SID-only decoding favors
    \(\stok{3}{13061}\), leading to the Bruno Fernandes video (a).  \textit{AGR}
    instead invokes the GSU tool.  Matching the prefix against the user's
    history yields five votes for \(\stok{3}{13842}\) versus one
    for \(\stok{3}{13061}\).  The returned observation
    \(\textcolor[HTML]{70C0C7}{o_{\mathrm G}\!:\!\stok{3}{13842}}\) redirects generation to the consumed Ronaldo
    video (d).

    \item \textcolor{CaseFreshnessColor}{\textbf{Temporal correction through
    Freshness.}}
    At \([\stok{1}{6421},\stok{2}{11480},\stok{3}{13706}]\), the original branch
    leads to an older July 3 match against Croatia (b).  The Freshness tool
    classifies the query as high-freshness and returns
    \(\textcolor[HTML]{F72D9F}{o_{\mathrm F}\!:\!\mathrm{High}}\),
    shifting generation toward \(\stok{4}{15244}\).  This retrieves the most
    recent report---Portugal's July 7 elimination by Spain (e).

    \item \textcolor{CaseIntentColor}{\textbf{Semantic disambiguation through
    Intent.}}
    At \([\stok{1}{6421},\stok{2}{11826},\stok{3}{14155}]\), the competing
    branch leads to a Portugal away-jersey video (c).  The Intent tool recognizes
    that the user seeks Sports information and returns
    \(\textcolor[HTML]{FF6B6B}{o_{\mathrm I}\!:\!\mathrm{Sports}}\).  This suppresses merchandise-oriented content and
    redirects generation toward \(\stok{4}{15376}\), retrieving a video about
    Portugal's 2026 World Cup squad (f).
\end{itemize}
% \red{
Together, these three independently tool-guided beam paths
% } 
produce a top-3 ranked set that covers the user's
personal preference, need for up-to-date results, and sports-intent content;
the first two returned videos were subsequently clicked.
\end{example}

% \paragraph{Additional cases.}
\centering
\includegraphics[width=\linewidth]{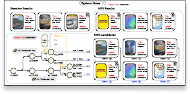}
\vspace{-9mm}
\caption{Case 2: Top-3 results for ``Typhoon News''}
\par\medskip
\includegraphics[width=\linewidth]{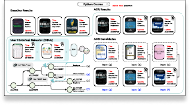}
\vspace{-9mm}
\caption{Case 3: Top-3 results for ``Python Course''}
\par\medskip
\includegraphics[width=\linewidth]{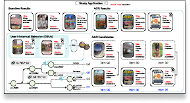}
\vspace{-9mm}
\caption{Case 4: Top-3 results for ``Study Application''}
\par\medskip
\includegraphics[width=\linewidth]{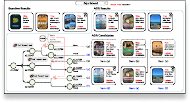}
\vspace{-9mm}
\caption{Case 5: Top-3 results for ``Jeju Island''}
\par\medskip
\includegraphics[width=\linewidth]{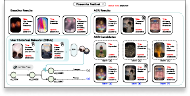}
\vspace{-9mm}
\caption{Case 6: Top-3 results for ``Fireworks Festival''}
\par\medskip
\raggedright

% \textbf{Takeaway.}
% The broader implication is that \textit{AGR} does not merely expose the
% retriever to more features; it determines which evidence is useful for each
% beam and when that evidence should be acquired.  This enables time- and
% context-adaptive interaction at different decoding branches, producing
% multi-channel-retrieval-like diversity within a unified generative retriever.

\end{tcolorbox}